\documentclass{vldb}
\usepackage{outlines}
\usepackage{graphicx}
\usepackage{amssymb}
\usepackage{multirow}
\usepackage{subfigure}
\usepackage{amsmath,thm-restate}
\usepackage{algorithmic}
\usepackage{algorithm}
\usepackage{subfigure}
\usepackage{color}
\usepackage[normalem]{ulem}
\usepackage{balance}
\newskip\subfigcapskip	\subfigcapskip	= 2ex

\addtocounter{MaxMatrixCols}{10}

\usepackage{url}

\toappear{}

\begin{document}

	\numberofauthors{2} 
	\author{
	\alignauthor 
	Chao Li \\
    \affaddr{University of Massachusetts \\Amherst, Massachusetts, USA}\\
    \affaddr{chaoli@cs.umass.edu}
	\alignauthor
	Gerome Miklau \\
    \affaddr{University of Massachusetts \\Amherst, Massachusetts, USA}\\
    \affaddr{miklau@cs.umass.edu}
	}	

\twocolumn
\setlength{\columnsep}{.7cm}
	
\title{An Adaptive Mechanism for Accurate \\ Query Answering under Differential Privacy}

\maketitle{}

\pagestyle{empty}


\abovedisplayskip = 3pt
\belowdisplayskip = 3pt
\subfigbottomskip=-8pt


\floatname{algorithm}{Program}
\renewcommand{\algorithmicrequire}{\textbf{Input:}}
\renewcommand{\algorithmicensure}{\textbf{Output:}}

\newcommand{\cell}{\phi}

\newcommand{\reals}{R}
\newcommand{\vol}{\textup{Vol}}
\newcommand{\convex}{\textup{Convex}}

\newcommand{\minerror}{\mbox{\sc MinError}}
\newcommand{\minsens}{\mbox{\sc MinSensitivity}}
\newcommand{\allrange}{\mbox{\sc AllRange}}
\newcommand{\allpred}{\mbox{\sc AllPredicate}}

\newcommand{\eqbydef}{\stackrel{\mathrm{def}}{=}}

\newcommand{\vect}[1]{\mathbf{#1}}
\newcommand{\sens}[1]{\Delta_{#1}}
\newcommand{\inv}[1]{{#1}^{-1}}
\newcommand{\ep}[1]{\inv{({#1}^t{#1})}}

\def\alg{\mathcal{K}}  
\def\LM{\mathcal{L}}	
\def\GM{\mathcal{G}}	
\def\MM{\mathcal{M}}	

\def\lbl{\mbox{LSA}}
\def\svdb{\mbox{\sc svdb}}
\def\cols{\mbox{cols}}

\def\tr{\mbox{trace}}
\def\var{\mbox{Var}}
\newcommand{\error}[2]{\mbox{\sc Error}_{#1}( #2 )}
\newcommand{\totalerror}[2]{\mbox{\sc TotalError}_{#1}( #2 )}
\newcommand{\maxerror}[2]{\mbox{\sc MaxError}_{#1}( #2 )}
\newcommand{\opterror}[1]{\mbox{\sc OptError}(#1)}
\newcommand{\apperror}[2]{\mbox{\small\sc EigError}_{#1}(#2)}
\newcommand{\optstrategy}[1]{\mbox{\small\sc OptStrat}(#1)}
\newcommand{\appstrategy}[2]{\mbox{\small\sc OptStrat}_{#1}(#2)}

\def\aa{\mathbb{A}}  
\def\bb{\mathbb{B}}  
\def\dsQ{\mathcal{Q}}

\def\plus{{\!+}}
\def\b{\vect{\tilde{b}}}  
\def\x{\vect{x}}  
\def\estx{\vect{\hat x}}
\def\y{\vect{y}}
\def\q{\vect{q}}  
\def\w{\vect{w}} 
\def\v{\vect{v}}  
\def\estw{\vect{\hat w}}
\def\estq{\vect{\hat q}}
\def\A{\vect{A}}
\def\T{\vect{T}}
\def\B{\vect{B}}
\def\Q{\vect{Q}}
\def\W{\vect{W}}
\def\M{\vect{M}}
\def\D{\vect{D}}
\def\P{\vect{P}}
\def\p{\vect{p}}
\def\I{\vect{I}}
\def\V{\vect{V}}
\def\H{\vect{H}}
\def\G{\vect{G}}
\def\R{\vect{R}}
\def\X{\vect{X}}
\def\Wav{\vect{Y}}
\def\lambdaB{\vect{\lambda}}
\def\LambdaB{\vect{\Lambda}}

\def\PM{\P_{\M}}
\def\DM{\D_{\M}}
\def\DS{\D_s}
\def\DSinv{\DS^{-1}}


\def\WW{\W}		
\def\Wbool{\W_{01}}		
\def\Wrang{\W_{R}}		
\def\Wunit{\W_{unit}}		

\def\real{\mathbb{R}}

\def\RR{\vect{R}}

\newcommand{\ff}[1]{#1}
\newcommand{\dif}[1]{\mathbf{\delta}_{#1}}


\newcommand{\cl}[1]{[[\emph{\color{blue}CL: #1}]]}
\newcommand{\mh}[1]{}
\newcommand{\gm}[1]{[[\emph{\color{red}GM: #1}]]}
\newcommand{\eat}[1]{}
\newcommand{\cut}[1]{}

\newcommand{\set}[1]{\{#1\}}   
\newcommand{\revision}[2]{#2}

\newtheorem{definition}{Definition}
\newtheorem{proposition}{Proposition}
\newtheorem{corollary}{Corollary}
\newtheorem{conjecture}{Conjecture}
\newtheorem{theorem}{Theorem}
\newtheorem{problem}{Problem}
\newtheorem{example}{Example}
\newtheorem{remark}{Remark}

\def\nbrs{nbrs}
\def\<{\langle}
\def\>{\rangle}

\def\qq{\tilde{q}}
\def\qbar{\overline{q}}

\def\Q{\mathbf{Q}}
\def\QQ{\mathbf{\tilde{Q}}}
\def\QC{\mathbf{\overline{Q}}}
\def\qq{\tilde{q}}
\def\qbar{\overline{q}}

\def\H{\mathbf{H}}
\def\HH{\mathbf{\tilde{H}}}
\def\HC{\mathbf{\overline{H}}}
\def\hh{\tilde{h}}
\def\hbar{\overline{h}}

\def\Lap{\mbox{Laplace}}
\def\Nor{\mbox{Normal}}
\def\cnt{c}
\def\cons{\gamma}
\def\db{I}

\def\hght{{\ell}}  
\def\hv{\hght(v)}
\def\wt{\alpha}
\def\root{r}

\newcommand{\frob}[1]{||#1||_f}
\newcommand{\E}{\mathbb{E}}
\newcommand{\Ldist}[3]{||#1 -#2||_{#3}}
\newcommand{\rank}{\textup{rank}}
\newcommand{\trace}{\textup{Trace}}

\newcommand{\Ltwo}[1]{||#1||_2}
\newcommand{\Lone}[1]{\left\Vert #1  \right\Vert_1}

\newcommand{\reffull}[1]{#1}

\newtheorem{lemma}{Lemma}

\newcommand{\one}[1]{\mathbb{I}_{#1}}
\def\U{\mathcal U}
\def\Z{succZ} 
\def\s{s}
\def\m{M}
\def\mm{\tilde{M}}


\begin{abstract}

We propose a novel mechanism for answering sets of counting queries under differential privacy.  Given a workload of counting queries, the mechanism automatically selects a different set of ``strategy'' queries to answer privately, using those answers to derive answers to the workload.  The main algorithm proposed in this paper approximates the optimal strategy for any workload of linear counting queries.  With no cost to the privacy guarantee, the mechanism improves significantly on prior approaches and achieves near-optimal error for many workloads, when applied under $(\epsilon, \delta)$-differential privacy.  The result is an adaptive mechanism which can help users achieve good utility without requiring that they reason carefully about the best formulation of their task.

\end{abstract}

\section{Introduction}

\begin{outline}[enumerate]

Differential privacy \cite{Dwork:2006Calibrating-Noise} guarantees that information released about participants in a data set will be virtually indistinguishable whether or not their personal data is included.  There are now many algorithms satisfying differential privacy~\cite{Dwork:2011A-firm-foundation}, however, when adopting differential privacy, users must reason carefully about alternative mechanisms and the formulation of their task.  Their choices may have a significant impact on the utility of the output, for the same level of privacy.  Even using the PINQ framework \cite{mcsherry2009privacy}, designed to aid uninitiated users in writing differentially-private programs, users can be faced with vastly different degrees of accuracy depending on how their task is expressed.

Further, there are few results showing that proposed algorithms are optimally accurate---that is, that they introduce the least possible distortion required to satisfy the privacy criterion.\footnote{For a single numerical query, the addition of appropriately-scaled discrete Laplace noise satisfies $\epsilon$-differential privacy and has been proven optimally accurate \cite{ghosh2009universally}. For workloads of multiple queries, optimally accurate mechanisms are not known.}  Thus, if the utility they achieve is unacceptable, users often do not know if better utility is possible with a different algorithm, or if their utility goals are fundamentally incompatible with differential privacy.  

In this work, we attempt to relieve the user of some of these difficulties by developing a mechanism that automatically adapts to the set of submitted queries and provides significantly improved utility over competing approaches.  We focus on batch query answering, in which a set of queries is answered at one time, in a single interaction with the private server.  We call the set of queries a {\em workload}, which we allow to be any collection of linear counting queries.  This general class of queries can be used to express histograms, marginals, data cubes, empirical cumulative distribution functions, common aggregation queries with grouping, and more. 

One of the motivations for considering batch query-ans\-wer\-ing of large workloads is to avoid the complications of online mechanisms in which a user must carefully manage their privacy budget, and, in addition, multiple users may be required to share a single privacy budget to avoid a breach of the privacy definition resulting from collusion.  It is therefore appealing to structure large workloads that contain the sufficient statistics of a data mining task, or which can simultaneously support the intended tasks of a group of users.  In fact, the output of our algorithms can often be treated as a synthetic data set, albeit one which is tailored specifically for accuracy on the queries in the given workload.

The standard approach for answering a workload of queries under $\epsilon$-differential privacy is the Laplace mechanism, which adds to each query a sample chosen independently at random from a Laplace distribution.  The noise distribution is scaled to the sensitivity of the workload: the maximum possible change to the query answers induced by the addition or removal of one tuple.  Large workloads often have high sensitivity, in which case the Laplace mechanism results in extremely noisy query answers because the noise added to {\em each} query in the workload is proportional to the sensitivity of the workload.  

Recently, a number of related approaches have been proposed which improve on the Laplace mechanism, sometimes allowing for low error where only unacceptably high error was possible before.  They each embody a basic (but perhaps counter-intuitive) principle: better results are possible when you {\em don't ask for what you want}.  

The earliest example of this approach focuses on workloads consisting of sets of k-way marginals, for which Barak et al.  answer a set of Fourier basis queries using the Laplace mechanism, and then derive the desired marginals~\cite{barak2007privacy}. For workloads consisting of all range-count queries over an ordered domain, two approaches have been proposed.  Xiao et al. \cite{xiao2010differential} first answer a set of wavelet basis queries, while Hay et al. \cite{Hay:2010Boosting-the-Accuracy} use a hierarchical set of counting queries which recursively decompose the domain.  For workloads consisting of sets of marginals, Ding et al.~\cite{Ding:2011fk} recently proposed a method for selecting an alternative set of marginals, from which the desired counts can be derived. 

These techniques can each be described in the framework of the recently-proposed matrix mechanism \cite{Li:2010Optimizing-Linear}.  Given a workload of queries, the matrix mechanism uses the Laplace mechanism to answer a set of {\em strategy} queries.  The answers to the strategy queries are then used to derive answers to the workload queries by finding a solution that minimizes squared error.  (The derivation by least squares is implicit in Barak \cite{barak2007privacy} and Xiao \cite{xiao2010differential}, but explicit in Hay \cite{Hay:2010Boosting-the-Accuracy} and Ding \cite{Ding:2011fk}).  In these terms, the four approaches described above can each be seen as providing a set of strategy queries suitable for a particular kind of workload.  Ultimately, the use of the strategy queries and the derivation process result in a more complex, non-independent noise distribution which can reduce error.

The matrix mechanism makes clear that nearly any set of strategy queries can be used in this manner to answer a workload.  Effective strategies have lower sensitivity than the workload, and are such that the workload queries can be concisely represented in terms of the strategy queries.  But the approach remains limited to specific strategies for range queries \cite{Hay:2010Boosting-the-Accuracy,xiao2010differential}, and approaches which provide only limited choices of strategies for marginals \cite{barak2007privacy,Ding:2011fk}.  

We continue this line of work in order to create a truly adaptive mechanism that can answer a wide range of workloads with low error.  The key to such a mechanism is {\em strategy selection}: the problem of computing the set of strategy queries that minimizes error for a given workload.  Unfortunately, exact solutions to the strategy selection problem are infeasible in practice~\cite{Li:2010Optimizing-Linear}. One of our main contributions is an approximation algorithm capable of efficiently computing a nearly optimal strategy in $O(n^4)$ time (where $n$ is the number of individual counting queries required to express the workload).  The result is a mechanism that adapts the noise distribution to the set of queries of interest, relieving the user of the burden of choosing among mechanisms or carefully analyzing their workload.

A few main insights underlie our contributions. First, we shift our focus to $(\epsilon,\delta)$-differential privacy, a modest relaxation of $\epsilon$-differential privacy.  The standard mechanism in this case is the Gaussian mechanism, which suffers the same limitations of the Laplace mechanism, and is also improved by the same approaches described above.  The important difference for our results is that sensitivity is measured using the $L_2$ metric (instead of $L_1$) which ultimately allows for better approximate solutions.\footnote{Our algorithm can also be adapted to $\epsilon$-differential privacy, but it is less efficient, appears to be less effective, and is significantly harder to analyze. (Please see Sec. \ref{sec:sub:l1}.)}  Second, inspired by the statistical problem of optimal experimental design~\cite{boyd2004convex,Pukelsheim93Optimal}, we formulate the strategy selection problem as a convex optimization problem which chooses $n$ coefficients to serve as weights for a fixed set of {\em design queries}. Third, we show that the eigenvectors of the workload (when represented in matrix form) capture the essential building blocks required for near-optimal strategies and are therefore a very effective choice for the design queries underlying the above optimization problem. 

Our adaptive mechanism advances the state-of-the-art in terms of accuracy, \revision{}{under both absolute and relative measures of error}: 

\begin{itemize} \itemsep 0in
\item[--] For workloads targeted by prior approaches, our algorithm automatically
computes strategies with uniformly lower error.  For marginals, our error can be reduced by as much as \revision{$19$}{$6.2$} times over Barak and \revision{$2.6$}{$3.2$} times over Ding.  For range queries, our error is reduced as much as \revision{$2.5$}{$2.6$} over Xiao and \revision{$3.5$}{$2.7$} times over Hay. 

\item[--] The power of our adaptive approach is most obvious when applying the mechanism to ad hoc workloads (which may result from specializing a larger workload to a given task, or by combining workloads from multiple users).  Error is reduced by as much as \revision{$30$}{$13$} times over alternative techniques. 

\item[--] Our algorithm has a provable approximation ratio and \revision{}{produces strategies with near optimal absolute error} for many workloads of interest.  We never witness an approximation rate greater than \revision{$1.6$}{$1.3$} times the optimal absolute error. For workloads of marginals, error rates consistently match the optimal achievable error rates.
\end{itemize}

Our mechanism is also significantly more general than prior work.  It can be applied to any workload of linear counting queries: a much larger class of queries than marginals or range queries.  In addition, the algorithm avoids a subtle limitation of some previous approaches \cite{Hay:2010Boosting-the-Accuracy,xiao2010differential,Ding:2011fk} in which achieving promised error rates depends on finding a proper representation for the workload.

Throughout the paper, all improvements to accuracy are made with {\em absolutely no cost to privacy}: accuracy is improved by constructing a better noise distribution satisfying the same privacy condition.
In addition, while strategy selection is the most comp\-utationally intensive part  of the mechanism, it only needs to be performed once for any workload, and need not be recomputed to re-run the mechanism on a new database instance.  Once the selected strategy is preprocessed, the complexity of executing the mechanism is no higher than applying the standard Laplace mechanism to the workload. 

The paper is organized as follows.  We review definitions and formally describe the matrix mechanism in Sec. \ref{sec:back}.  Our algorithm is presented in Sec. \ref{sec:alg}, along with a theoretical analysis that establishes the approximation rate and other properties.  In Sec. \ref{sec:eff} we propose performance optimizations which  significantly improve computation time with minimal impact on solution quality.  In Sec. \ref{sec:exp}, we  evaluate \revision{}{both absolute and relative} error rates of our mechanism on a range of workloads. \revision{}{We discuss related work and conclude in Sec. \ref{sec:related} and Sec. \ref{sec:conclusion}.}

\end{outline}

\section{Background} \label{sec:back}

In this section we describe our data model and privacy conditions. We also review the fundamentals of the matrix mechanism, including error measurement and the problem of strategy selection. \revision{}{Throughout the paper, we use the notation of linear algebra and employ standard techniques of matrix analysis. For a matrix $\A$, $\A^T$ is its transpose and $\tr(\A)$ is the sum of values on the main diagonal. If $\A$ is a square matrix with full rank,  $\A^{-1}$ denotes its inverse. We use $diag(c_1, \dots c_n)$ to indicate an $n \times n$ diagonal matrix with scalars $c_i$ on the diagonal.}
\subsection{\hspace*{-9pt}Data Model, Linear Queries, and Workloads}

The workloads considered in this paper consist of counting queries over a single relation.  Let the database $I$ be an instance of a single-relation schema $R(\mathbb{A})$, with attributes $\mathbb{A}=\{A_1, A_2, \ldots, A_k\}$.  The crossproduct of the attribute domains, written $dom(\mathbb{A})$, is the set of all possible tuples that may occur in $I$.  

In order to express our queries, we first transform the instance $I$ into a {\em data vector} $\x$ of cell counts.  We may choose to fully represent instance $I$ by defining the vector $\x$ with one cell for every element of $dom(\mathbb{A})$.  Then $\x$ is a bit vector of size $|dom(\mathbb{A})|$ with nonzero counts for each tuple present in $I$.  This is often inefficient (the size of the $\x$ vector is the product of the attribute domain sizes) and ineffective (the base counts are typically too small to be estimated accurately under the privacy condition). A common way to form a vector of base counts over larger cells is to partition each $dom(A_i)$ into $d_i$ regions, which could correspond to ranges over an ordered domain or individual elements (or sets of elements) in a categorical domain.  Then the individual cells are defined by taking the cross-product of the regions in each attribute.  
The choice of cells in the data vector is ultimately determined by the workload queries that need to be expressed.  

To formally define the data vector we associate, with each element $x_i$ of $\x$, a Boolean {\em cell condition} $\cell_i$, which evaluates to True or False for any tuple in $dom(\mathbb{A})$.  We always require that the cell conditions be pairwise unsatisfiable: any tuple in $dom(\mathbb{A})$ will satisfy exactly one cell condition.  Then $x_i$ is defined to be the count of the tuples from $I$ which satisfy $\cell_i$. 
\begin{definition}[Data vector]
Given an ordered list of cell conditions $\cell_1, \cell_2 \dots \cell_n$ the data vector $\x$ is a length-$n$ column vector defined by $n$ positive integral counts $x_i = |\{t \in I \:|\: \cell_i(t) \mbox{ is True}\}|$.  
\end{definition}
\vspace*{-1pt}
In the sequel, the length of $\x$ is a key parameter, always denoted by $n$.
\begin{example}  Consider the relational schema $R (name,\\ gradyear, gender, gpa)$ describing students.  If we wish to form queries only over $gender$ (Male or Female), and $gpa$ ranges $[1.0,2.0)$, $[2.0,3.0)$, $[3.0,3.5)$, $[3.5,4.0)$, then we can define the 8 cell conditions enumerated in Fig.~\ref{tbl:one}(a).
\end{example}
\vspace*{-1pt}

A linear query computes a specified linear combination of the elements of the data vector $\x$. 
\begin{definition}[Linear query]
A {\em linear query} is a \\length-$n$ row vector $\q=[q_1 \dots q_n]$ with each $q_i \in \mathbb{R}$.  
The answer to a linear query $\q$ on $\x$ is the vector product $\q\x = q_1x_1 + \dots + q_nx_n$.
\end{definition}
\revision{}{In addition to basic predicate counting queries, other aggregates like sum and  average, as well as group-by queries, can be expressed as linear counting queries.}  A {\em workload} is a set of linear queries.  A workload is represented as a matrix, where each row is a single linear counting query. 

\begin{definition}[Query matrix]
A {\em query matrix} is a collection of $m$ linear queries, arranged by rows to form an $m \times n$ matrix.
\end{definition}
If $\W$ is an $m \times n$ query matrix, the query answer for $\W$ is a length $m$ column vector of query results, which can be computed as the matrix product $\W \x$.  Note that cell condition $\cell_i$ defines the meaning of the $i^{th}$ position of $\x$, and accordingly, it determines the meaning of the $i^{th}$ column of a query matrix.  

\begin{example}   
Fig.~\ref{tbl:one}(b) shows a query matrix representing a workload of 8 linear queries.  Fig.~\ref{tbl:one}(c) describes the meaning of the queries w.r.t. the cell conditions in Fig.~\ref{tbl:one}(a).
\end{example}

Note that the data analyst should include in the workload {\em all} queries of interest, even if some queries could be computed from others in the workload.  In the absence of noise introduced by the privacy mechanism, it might be reasonable for the analyst to request answers to a small set of counting queries, from which other queries of interest could be computed.  (E.g., it would be sufficient to recover $\x$ itself by choosing the workload defined by the identity matrix.)  But because the analyst will receive private, noisy estimates to the workload queries, the error of queries computed from their combination is often increased.  Our adaptive mechanism is designed to optimize error across the entire set of desired queries, so all queries should be included.  As a concrete example, in Fig.~\ref{tbl:one}(b), $\q_3$ can be computed as $(\q_1 - \q_2)$ but is nevertheless included in the workload.  

We introduce terminology for a few common workloads used throughout the paper. The relevant properties of workloads are reflected by their matrix representation, so we often drop explicit mention of the schema and attributes involved and focus simply on the number of distinct attributes and the number of disjoint buckets for each attribute, assuming that cells are formed uniformly in the manner described above. 

We consider {\em predicate queries}, {\em range queries} and  {\em $k$-way marginal queries}. In addition, since each $k$-way marginal query covers a single value on the margin, one may need to sum answers to multiple marginal queries in order to answer any range query on the margin. When the answers to the marginal queries have noise, summing introduces too much accumulated noise. Therefore, in this paper, we also consider {\em $k$-way range marginal queries}, each of which aggregates multiple $k$-way marginal queries so as to cover a range on the margin.
%

\begin{example}
Of the queries in Fig.~\ref{tbl:one}, the first seven\\ are range queries (and therefore predicate queries as well). $\q_1 \dots \q_5$ are one-way range marginal queries, in which $\q_1$, $\q_2$, $\q_3$ are one-way range marginal queries over gender and $\q_1, \q_4, \q_5$ are one-way range marginal queries over gpa; $\q_2,\q_3$ are also one-way marginal queries.
\end{example}

We often consider large workloads consisting of {\em all} queries of a given type, such as ``all predicate'', ``all range'', ``all $k$-way range marginal'', ``all $k$-way marginal'' and ``all marginal'' (the union of all $k$-way marginals for $0\leq k\leq m$). Notice there is no workload of ``all range marginal'' since it is equivalent to ``all range''.  Later we will also consider {\em ad hoc} workloads consisting of arbitrary subsets of each of these types of queries and their combinations.  In practice such workloads are important because they may arise from combining queries of interest to multiple users, or from specializing a general workload to a more specific task, to improve error.

\begin{figure*} 
\vspace{1ex}
\centering
\subfigure[Cell conditions $\Phi$]{
\small
\begin{tabular}{l}
$\cell_1: gpa\in[1.0,2.0) \wedge gender=M$\\
$\cell_2: gpa\in[2.0,3.0) \wedge gender=M$ \\
$\cell_3: gpa\in[3.0,3.5) \wedge gender=M$ \\
$\cell_4: gpa\in[3.5,4.0) \wedge gender=M$ \\
$\cell_5: gpa\in[1.0,2.0) \wedge gender=F$ \\
$\cell_6: gpa\in[2.0,3.0) \wedge gender=F$ \\
$\cell_7: gpa\in[3.0,3.5) \wedge gender=F$ \\
$\cell_8: gpa\in[3.5,4.0) \wedge gender=F$ \\
\end{tabular}
}
\quad
\subfigure[A query workload $\W$]{
\small
$\begin{bmatrix}
1 & 1 & 1 & 1 & 1 & 1 & 1 & 1 \\
1 & 1 & 1 & 1 & 0 & 0 & 0 & 0 \\
0 & 0 & 0 & 0 & 1 & 1 & 1 & 1 \\
1 & 1 & 0 & 0 & 1 & 1 & 0 & 0 \\
0 & 0 & 1 & 1 & 0 & 0 & 1 & 1 \\
0 & 0 & 0 & 0 & 0 & 0 & 1 & 1 \\
1 & 1 & 0 & 0 & 0 & 0 & 0 & 0 \\
1 & 1 & 1 & 1 & \mbox{-}1 & \mbox{-}1 & \mbox{-}1 & \mbox{-}1 \\
\end{bmatrix}$
}
\quad
\subfigure[Counting queries defined by rows of $\W$]{ \small
\begin{tabular}{l}
$\q_1$: all students; \\
$\q_2$: female students;\\
$\q_3$: male students;\\
$\q_4$: students with $gpa < 3.0$;\\
$\q_5$: students with $gpa \ge 3.0$;\\
$\q_6$: female students with $gpa \ge 3.0$;\\
$\q_7$: male students with $gpa < 3.0$;\\
$\q_8$: difference between male and female students.\\
\end{tabular}
}

\vspace*{3pt}
\caption{\label{tbl:one} For schema $R=(name, gradyear, gender, gpa)$, (a) shows 8 cell conditions on attributes $gender$ and $gpa$.  The database vector $\x$ (not shown) will accordingly consist of 8 counts; (b) shows a sample workload matrix $\W$ consisting of 8 queries, each described in (c).}
\vspace*{-12pt}
\end{figure*} 


\subsection{Differential Privacy and Gaussian Noise}

Standard $\epsilon$-differential privacy \cite{Dwork:2006Calibrating-Noise} places a bound (controlled by $\epsilon$) on the difference in the probability of query answers for any two {\em neighboring} databases.  For database instance $\db$, we denote by $\nbrs(\db)$ the set of databases differing from $\db$ in at most one record. 
Approximate differential privacy~\cite{Dwork:2006Our-Data-Ourselves:,McSherry:2009fk}, is a modest relaxation in which the $\epsilon$ bound on query answer probabilities may be violated with small probability (controlled by $\delta$).

\begin{definition}\hspace*{-2pt}{\sc(Approximate Differential Privacy)} A randomized algorithm $\alg$ is $(\epsilon,\delta)$-differentially private if for any instance $I$, any $I' \in \nbrs(I)$, and any subset of outputs $S \subseteq Range(\alg)$, the following holds:
\[
Pr[ \alg(I) \in S] \leq \exp(\epsilon) \times Pr[ \alg(I') \in S] +\delta
\]		
	\end{definition}
When $\delta=0$, the condition is standard $\epsilon$-differential privacy.

Both definitions can be satisfied by adding random noise to query answers.  The magnitude of the required noise is determined by the {\em sensitivity} of a set of queries: the maximum change in a vector of query answers over any two neighboring databases.  But the two privacy definitions differ in the measurement of sensitivity and in their noise distributions.  Standard differential privacy can be achieved by adding Laplace noise calibrated to the $L_1$ sensitivity of the queries \cite{Dwork:2006Calibrating-Noise}. Approximate differential privacy can be achieved by adding Gaussian noise calibrated to the $L_2$ sensitivity of the queries \cite{Dwork:2006Our-Data-Ourselves:,McSherry:2009fk}.  
Our main results focus on approximate differential privacy, but we discuss extensions to standard differential privacy in Sec~\ref{sec:sub:l1}.

Our query workloads are represented as matrices, so we express the sensitivity of a workload as a matrix norm.  Because neighboring databases $\db$ and $\db'$ differ in exactly one tuple, and because cell conditions are disjoint, it follows that the corresponding vectors $\x$ and $\x'$ differ in exactly one component, by exactly one, in which case we write  $\x' \in \nbrs(\x)$.  The $L_2$ sensitivity of $\W$ is equal to the maximum $L_2$ norm of the columns of $\W$. Below, $\cols(\W)$ is the set of column vectors $W_i$ of $\W$.

\begin{proposition}[$L_2$ Query matrix sensitivity]
The $L_2$ sensitivity of a query matrix $\W$ is denoted $\Ltwo{\W}$, defined as follows:
\begin{eqnarray*}
\Ltwo{\W} & \eqbydef & \max_{\x' \in \nbrs(\x)} \Ltwo{\W\x - \W\x'} 
			= \max_{W_i \in \cols(\W)} \Ltwo{W_i}
\end{eqnarray*}
\end{proposition}
For example, for $\W$ in Fig.~\ref{tbl:one}(b), we have $\Ltwo{\W}=\sqrt{5}$.

The classic differentially private mechanism adds independent noise calibrated to the sensitivity of a query workload.  We use $\Nor(\sigma)^m$ to denote a column vector consisting of $m$ independent samples drawn from a Gaussian distribution with mean $0$ and scale $\sigma$.

\begin{proposition}{\sc (Gaussian mechanism \cite{Dwork:2006Our-Data-Ourselves:, McSherry:2009fk})}\label{thm:l2diffpriv}
Given an $m \times n$ query matrix $\W$, the randomized algorithm $\GM$ that outputs the following vector is $(\epsilon,\delta)$-differentially private:
$$\GM(\W,\x) = \W\x + \Nor(\sigma)^m$$ 
where $\sigma=\Ltwo{\W}\sqrt{2\ln(2/\delta)}/\epsilon$
\end{proposition}

Recall that $\W\x$ is a vector of the true answers to each query in $\W$.  The algorithm above adds independent Gaussian noise (scaled by $\epsilon$, $\delta$, and the sensitivity of $\W$) to each query answer.  Thus $\GM(\W,\x)$ is a length-$m$ column vector containing a noisy answer for each linear query in $\W$.

\subsection{The Matrix Mechanism}

The matrix mechanism has a form similar to the Gaussian mechanism but adds a more complex noise vector.  It uses a strategy matrix, $\A$, to construct this vector.

\begin{proposition}{\sc ($(\epsilon,\delta)$-Matrix Mechanism \cite{Li:2010Optimizing-Linear})} \label{def:m-mech} 
Given an $m \times n$ query matrix $\W$, and assuming $\A$ is a full rank $p \times n$ strategy matrix, the randomized algorithm $\MM_\A$ that outputs the following vector is $(\epsilon,\delta)$-differentially private:
\begin{eqnarray*}
\MM_\A(\W,\x) &=& \W\x + \W \A^\plus \Nor(\sigma)^m.
\end{eqnarray*}
where $\sigma=\Ltwo{\A}\sqrt{2\ln(2/\delta)}/\epsilon$
\end{proposition}

Here $\A^\plus$ is the pseudo-inverse of $\A$: $\A^\plus = \inv{(\A^T\A)}\A^T$; if $\A$ is a square matrix, then $\A^\plus$ is just the inverse of $\A$.  The intuitive justification for this mechanism is that it is equivalent to the following three-step process: (1) the queries in the strategy are submitted to the Gaussian mechanism; (2) an estimate $\estx$ for $\x$ is derived by computing the $\estx$ that minimizes the squared sum of errors (this step consists of standard linear regression and requires that $\A$ be full rank to ensure a unique solution); (3) noisy answers to the workload queries are then computed as $\W\estx$.  The answers to $\W$ derived in step (3) are always consistent because they are computed from a single noisy version of the cell counts, $\estx$.

Like the Gaussian mechanism, the matrix mechanism computes the true answer vector $\W\x$ and adds noise to each component.  But a key difference is that the scale of the Gaussian noise is {\em calibrated to the sensitivity of the strategy matrix $\A$, not that of the workload}.  In addition, the noise added to query answers is no longer independent, because the vector of independent Gaussian samples is transformed by the matrix $\W\A^\plus$.  




\begin{figure}[t]
\centering 
\begin{tabular}{cc} 
Identity & Wavelet \vspace{2ex} \\
\small $\left[\begin{smallmatrix}
1 & 0 & 0 & 0 & 0 & 0 & 0 & 0 \\
0 & 1 & 0 & 0 & 0 & 0 & 0 & 0 \\
0 & 0 & 1 & 0 & 0 & 0 & 0 & 0 \\
0 & 0 & 0 & 1 & 0 & 0 & 0 & 0 \\
0 & 0 & 0 & 0 & 1 & 0 & 0 & 0 \\
0 & 0 & 0 & 0 & 0 & 1 & 0 & 0 \\
0 & 0 & 0 & 0 & 0 & 0 & 1 & 0 \\
0 & 0 & 0 & 0 & 0 & 0 & 0 & 1 \\
\end{smallmatrix}\right]$ 
\hfill &  
$\left[\begin{smallmatrix}
1 & 1 & 1 & 1 & 1 & 1 & 1 & 1 \\
1 & 1 & 1 & 1 & \mbox{-}1  & \mbox{-}1  & \mbox{-}1  & \mbox{-}1  \\
1 & 1 & \mbox{-}1  & \mbox{-}1  & 0 & 0 & 0 & 0 \\
0 & 0 & 0 & 0 & 1 & 1 & \mbox{-}1  & \mbox{-}1  \\
1 & \mbox{-}1  & 0 & 0 & 0 & 0 & 0 & 0 \\
0 & 0 & 1 & \mbox{-}1  & 0 & 0 & 0 & 0 \\
0 & 0 & 0 & 0 & 1 & \mbox{-}1  & 0 & 0 \\
0 & 0 & 0 & 0 & 0 & 0 & 1 & \mbox{-}1  \\
\end{smallmatrix}\right]$ \vspace{2ex}
\\ 
\multicolumn{2}{c}{Adaptive Algorithm Output} \vspace{2ex} \\
\multicolumn{2}{c}{
$\left[\begin{smallmatrix}
0 & 0 & \mbox{-}.1 & .1 & .19 & \mbox{-}.19 & 0 & 0 \\
0 & 0 & .19 & \mbox{-}.19 & .1 & \mbox{-}.1 & 0 & 0 \\
\mbox{-}.3 & \mbox{-}.3 & .33 & .33 & .33 & .33 & \mbox{-}.3 & \mbox{-}.3 \\
.47 & .47 & \mbox{-}.5 & \mbox{-}.5 & .5 & .5 & \mbox{-}.47 & \mbox{-}.47 \\
.56 & .56 & .53 & .53 & \mbox{-}.53 & \mbox{-}.53 & \mbox{-}.56 & \mbox{-}.56 \\
.62 & .62 & .57 & .57 & .57 & .57 & .62 & .62 \\
\end{smallmatrix}\right]$
}  \\ 
\end{tabular}
\vspace*{-.5ex}
\caption{\label{fig:strat} Alternative strategy matrices that can be used to answer  workload $\W$ from Fig \ref{tbl:one}(b).  The \revision{total error of $\W$}{root mean square error of answering $\W$} when using the identity, wavelet, and adapted strategies is \revision{$36.0$, $21.0$, and $15.5$,}{$45.36$, $34.62$ and $29.79$,} respectively. }
\vspace*{-10pt}

\end{figure}

\begin{example}  
The three strategy matrices in Fig.~\ref{fig:strat} can be used by the matrix mechanism to answer the workload $\W$ in Fig.~\ref{tbl:one}(b), with differing results.  The first strategy is the identity matrix, the second is the Haar wavelet strategy, and the third is the output of the algorithm proposed in Sec \ref{sec:alg}. \revision{If the workload itself is used as the strategy, the total error of $\W$}{With $\epsilon=0.5$ and $\delta=0.0001$, if the workload itself is used as the strategy, the root mean square error of answering $\W$} is \revision{$40.0$}{$47.78$}.  The \revision{total error}{root mean square error} using the identity, wavelet, and adaptive strategies is \revision{$36.0$, $21.0$, and $15.5$,}{$45.36$, $34.62$ and $29.79$,} respectively.  It is possible to prove that no strategy can answer $\W$ with error less than \revision{$14.93$}{$29.18$}, so the algorithm is finding a nearly optimal strategy for this workload. 

Intuitively, by using the identity strategy, we get noisy estimates of each cell count using the Gaussian mechanism, and then use those estimates to compute the workload queries.  This strategy performs poorly for workload queries that sum many base counts because the variance of the independent noise increases additively.  The wavelet addresses this limitation by allowing large range queries to be estimated by combining the answers to just a few of the wavelet strategy queries.  It offers a dramatic improvement over the identity strategy for workload consisting of all range queries.  However, the wavelet is not necessarily appropriate for every workload. Our algorithm produces a strategy customized to $\W$, allowing for reduced error. \revision{(The error rates above assume for simplicity that the privacy coefficient $P(\epsilon,\delta)$ is 1.)}{}

\end{example}

\subsection{Optimal Error for the Matrix Mechanism} \label{sec:sub:error}

We measure the accuracy of a noisy query answer using \revision{mean squared error}{root mean square error.}  For a workload of queries, the error is defined as the \revision{sum}{root mean square error} of \revision{individual query errors}{the vector of answers, which we refer to simply as {\em workload error} in the remainder of the paper.}
 
\begin{definition}[Query and Workload Error]\label{def:errors} 
Let \\$\estw$ be the estimate for query $\w$ under the matrix mechanism using query strategy $\A$.  That is, $\estw=\MM_\A(\w,\x)$.  The \revision{mean squared error}{query error} of the estimate for $\w$ using strategy $\A$ is:
\revision{$$\error{\A}{\w} \eqbydef \E[ ( \w\x - \estw)^2 ].$$}{$$\error{\A}{\w} \eqbydef \sqrt{\E[ ( \w\x - \estw)^2 ]}.$$ } 
Given a workload $\W$\revision{, the  \em total mean squared error}{ consisting of $m$ queries, the workload error} of answering $\W$ using strategy $\A$ is:  
\revision{$\error{\A}{\W}\eqbydef$\\$\sum_{\w_i \in \W} \error{\A}{\w_i}$.}{\[\error{\A}{\W}\eqbydef\\\sqrt{\frac{1}{m}\sum_{\w_i \in \W} \error{\A}{\w_i}^2}.\]}
\vspace*{-2ex}
\end{definition}

The query answers returned by the matrix mechanism are linear combinations of noisy strategy query answers to which independent Gaussian noise has been added.  Thus, as the following proposition shows, we can directly compute the error for any linear query $\w$ or workload $\W$ as a function of $\epsilon,\delta$, and $\A$:


\begin{proposition}{\sc (\revision{Total Error}{Workload Error})} \label{prop:totalerror}
Given a workload $\W$, the \revision{total error}{error} of answering $\W$ using the $(\epsilon,\delta)$ matrix mechanism with query strategy $\A$ is:
\revision{
\begin{equation}\label{eqn:totalerror}
 \error\A{\W} = P(\epsilon, \delta)|| \A ||_2^2 \;\tr (\W^T\W(\A^T\A)^{-1})
 \end{equation}
 }
 {
\begin{equation}\label{eqn:totalerror}
 \error\A{\W} = ||\A||_2\sqrt{P(\epsilon, \delta)\;\tr (\W^T\W(\A^T\A)^{-1})}
 \end{equation}
 }
where $P(\epsilon, \delta)=\frac{2\log(2/\delta)}{\epsilon^2}$.
\end{proposition}
\eat{According to proposition~\ref{prop:totalerror}, when the privacy parameters $\epsilon$ and $\delta$ are chosen, the total error of using strategy $\A$ to answer workload $\W$ is explicitly determined by 
\[|| \A ||_2^2 \tr (\W^T\W(\A^T\A)^{-1}).\]
To make the representation easier, in the rest of this paper, we always assume $P(\epsilon, \delta)=1$ so that
\[ \totalerror\A{\W} = || \A ||_2^2\tr (\W^T\W(\A^T\A)^{-1}).\]}
%

To build a mechanism that adapts to a given workload $\W$, our goal is to select a strategy $\A$ to minimize the above formula.  The optimal strategy for a workload $\W$ is defined to be one that minimizes the \revision{total error:}{workload error:}

\begin{problem}{\sc (Optimal Strategy Selection)}  \label{prob:mintotal}
Given a workload $\W$, find a query strategy $\A_0$ such that:
\begin{equation}\label{eqn:mintotalerror}
	\error{\A_0}\W = {\min}_\A\error\A\W.
\end{equation}
\end{problem}

We denote the problem of computing an optimal strategy matrix as $\optstrategy{\W}$ and the \revision{}{workload error} under this strategy as $\opterror{\W}$.  It is possible to compute an exact solution to $\optstrategy{\W}$ by representing it as a convex optimization problem~\cite{Li:2010Optimizing-Linear}.  However, encoding the necessary constraints results in a problem with a large number of variables and optimization takes $O(n^8)$ time with standard solvers, making it infeasible for practical applications.  One of our main goals is to efficiently find approximately optimal strategy matrices, for any provided workload. 

\revision{}{We emphasize that the algorithms in this paper optimize the workload error, an {\em absolute} measure of error. The solution to this optimization problem depends on the workload alone, not on the input database. (This is evident from the fact that $\x$, the vector of database counts, does not appear in Eq. (\ref{eqn:totalerror}) above.)  We also consider {\em relative} error in the experimental evaluation, which inherently depends on the input database.  We show that low relative error for a workload $\W$ can be achieved by optimizing (absolute) error of a workload whose rows have been scaled in a straightforward way.}


%
%
%
%

\section{An Algorithm for Efficient\\ Strategy Selection} \label{sec:alg}

In this section we present an approximation algorithm for the strategy selection problem, prove its approximation rate and other properties, and discuss adapting the algorithm to $\epsilon$-differential privacy. 


\subsection{Optimal Query Weighting}

The main difficulty in solving $\optstrategy{\W}$ is computing (subject to complex constraints) all $n^2$ entries of a strategy matrix.  To simplify the problem, we take inspiration from the related problem of {\em optimal experimental design} \cite{Pukelsheim93Optimal}.

Consider a scientist who wishes to estimate the value of $n$ unknown variables as accurately as possible. The variables cannot be observed directly, but only by running one or more of a fixed set of feasible experiments, each of which returns a linear combination of the variables.  The experiments suffer from observational error, but those errors are assumed independent, and it follows that the least square method can be used to estimate the unknown variables once the results of the experiments are collected. Each experiment has an associated cost (which may represent time, effort, or financial expense) and the scientist has a fixed budget.  The optimal experimental design is the subset (or weighted subset) of feasible experiments offering the best estimate of the unknown variables and with a cost less than the budget constraint.  


There is an immediate analogy to the problem of strategy selection: our strategy queries are like experiments that provide partial information about the unknown data vector $\x$, and the final result will be computed using the least square method. However, in our setting, we are permitted to ask any query, with a cost (arising from the increase in sensitivity) which impacts the added noise.  In addition, our goal is to minimize the \revision{sum of variances of the given workload queries}{workload error}, while experimental design always minimizes the error of the individual variables (i.e. the error metric in experimental design is equivalent to our problem only if $\W$ is the identity matrix).

Despite these important differences, we adopt from experimental design the idea to limit the selection of our strategy to weighted combinations of a set of {\em design queries} that are fixed ahead of time.  Naturally, design queries with a weight of zero are omitted.  For a set of design queries $\dsQ$, the following problem, denoted $\appstrategy{\dsQ}{\W}$, selects the set of weights which minimizes the \revision{total error}{workload error} for $\W$.

\begin{problem}[Approximate Strategy Selection]  \label{prob:expdesign}
$ $\\Let $\W$ be a workload and $\dsQ=\{\q_1, \dots \q_k\}$ the design queries. For weights $\LambdaB=(\lambda_1 \dots \lambda_k) \in \mathbb{R}^k$, let matrix $\A_{\LambdaB, \dsQ}=$\\ $[\lambda_1\q_, \ldots, \lambda_k\q_k]^T$. Choose weights $\LambdaB_0\in\mathbb{R}^k$ such that:
\begin{equation}\label{eqn:approxtotalerror}
	\error{\A_{\LambdaB_0, \dsQ}}\W = {\min}_{\LambdaB\in\mathbb{R}^k}\error{\A_{\LambdaB, \dsQ}}\W.
\end{equation}
\end{problem}

The solution to this problem only approximates the truly optimal strategy since it is limited to selecting a strategy that is a weighted combination of the design queries.  But $\appstrategy{\dsQ}{\W}$ can be computed much more efficiently than $\optstrategy{\W}$. To do so, we describe $\appstrategy{\dsQ}{\W}$ as a semi-definite program~\cite{boyd2004convex}, a special form of convex optimization in which a linear objective function is minimized over the cone of positive semidefinite matrices.  Below, $\circ$ is the Hadamard (entry-wise) product of two matrices, and for \revision{}{symmetric} matrix $\Q$, $\Q\succeq 0$ \revision{means}{denotes} that $\Q$ is positive semidefinite\revision{.}{, which means $\x^T\Q\x\geq 0$ for any vector $\x$.} 
\begin{algorithm}[ht]
\small
\vspace{-1ex}
\begin{align*}
\mbox{\textbf{Given: }} & c_1,\ldots,c_n, \,\dsQ=[\q_1, \ldots, \q_n].\\
\mbox{\textbf{Choose: }} & u_1,\ldots,u_n,\,v_1,\ldots,v_n.\\
\mbox{\textbf{Mimimize: }} & c_1v_1+\ldots+c_nv_n.\\
\mbox{\textbf{Subject to: }} & \left[\begin{array}{cc}u_i & 1\\1 & v_i\end{array}\right]\succeq 0, \quad i=1,\ldots, n.\\
& (\dsQ\circ \dsQ)^T\mathbf{u}\leq \mathbf{1}.
\end{align*}
\vspace{-2ex}
\caption{Optimal Query Weighting}
\label{prog:expdesign}
\end{algorithm}
\vspace*{-3ex}

\begin{theorem} \label{thm:sdp}
Given a workload $\W$ and a set of design queries $\dsQ=\{\q_1, \dots \q_n\}$, let $c_1,\ldots, c_n$ be the squared $L_2$ norms of the columns of matrix $\W\dsQ^{+}$.  If the output of Program \ref{prog:expdesign} is $u_1,\ldots, u_n$ then setting $\LambdaB=\{\sqrt{u_1} \dots \sqrt{u_n}\}$ achieves $\appstrategy{\dsQ}{\W}$.  
\end{theorem}
\revision{
\begin{proof}
(Sketch) To solve Problem~\ref{prob:expdesign}, notice that applying a scalar to $\lambda_1,\ldots,\lambda_n$ will not change the value of $\error{\A(\LambdaB,\Q)}{\W}$.  Thus we can constrain the sensitivity of the strategy to be $1$.  Then the problem is equivalent to minimizing $c_1/\lambda_1^2+\ldots+c_n/\lambda_n^2$ with the constraint that the sensitivity of the strategy is 1.  In Program~\ref{prog:expdesign}, the larger $u_i$ leads to smaller $v_i$ so that smaller minimization goal. Thus the semidefinite constraints guarantee that $v_i=1/u_i$ and the inequality constraints require the sensitivity to be 1 for any optimal solution. 
\end{proof}
}{}
Algorithms for efficiently solving semidefinite programs have received considerable attention recently \cite{boyd2004convex}.  Using standard algorithms, Program~\ref{prog:expdesign} can be solved in $O(n|\dsQ|^3)$ time.  Recall that the complexity of computing $\optstrategy{\W}$ is $O(n^8)$.  Thus, Program~\ref{prog:expdesign} offers an efficiency improvement as long as $|\dsQ|=O(n^2)$.  This provides a target size for selecting the design set, which we turn to next.

\subsection{Choosing the Design Queries}\label{sec:alg:choose}
The potential of the above approach depends on finding a set of design queries, $\dsQ$, that is concise (containing no more than $n^2$, and preferably $n$, queries) and also expressive (so that near-optimal solutions can be expressed as weighted combinations of its elements).

One straightforward idea is to adopt as the design queries one of the proposed strategy matrices from prior work.  These are good strategy matrices for specific workloads such as the set of all range queries (wavelet or hierarchical strategy) or sets of low order marginals (the Fourier strategy).  Choosing one of these for $\dsQ$ would guarantee that $\appstrategy{\dsQ}{\W}$ produces a solution that improves upon the error of using that strategy.  Unfortunately these strategies are not sufficiently expressive for workloads very different from their target workloads.

Another possibility is to use the workload itself as the set of design queries, but there are two difficulties with this. First, there is no guarantee that a workload includes within it the components from which a high quality strategy may be formed, especially if the workload only contains a small set of queries.  The workloads of all range and all predicate queries are in fact sufficiently expressive (e.g. both the hierarchical strategy and a strategy equivalent to wavelet can be constructed by applying weights to the set of all range queries).  But this leads to the second issue: these workloads, and others that serve important applications, are too large and fail to meet our conciseness requirement. 

To avoid these pitfalls, we will derive the design set from the given workload $\W$ by applying tools of spectral analysis.  Intuitively this is a good choice because the eigenvectors of a matrix often capture its most important properties.  We will also show in the next section that this choice aids in the theoretical analysis of the approximation ratio because it allows us to relate the output of 
$\appstrategy{\dsQ}{\W}$ to a lower bound on error that is a function of the workload eigenvalues.

Recall that the key part of the expression for Eqn.~(\ref{eqn:totalerror}) in Prop. \ref{prop:totalerror} is $\tr (\W^T\W(\A^T\A)^{-1})
$, and notice that the workload occurs only in the form of $\W^T\W$.  It follows that there are many workloads with equivalent \revision{total error }{error} because it is easy to construct a matrix $\W_0$ such that $\W_0^T\W_0=\W^T\W$ by letting $\W_0=\Q\W$ for any orthogonal matrix $\Q$.  This suggests that, as far as \revision{total error}{workload error}  under the matrix mechanism is concerned, the essential properties of the workload are reflected by $\W^T\W$.  This motivates the following definition of eigen-queries of a workload, which we will use as our design set.
\begin{definition}[Eigen-queries of a workload]$ $\\
Given a workload $\W$, consider the eigen-decomposition of $\W^T\W$ into $\W^T\W=\Q^T\D\Q$, where $\Q$ is an orthogonal matrix and $\D$ is a diagonal matrix. \eat{whose diagonal entries are $\sigma_1, \ldots, \sigma_n$. } The \textbf{eigen-queries} of $\W$ are the rows of $\Q$ (i.e. the eigenvectors of $\W^T\W$). 
\end{definition}

Choosing the eigen-queries of $\W$ as the design set meets our conciseness requirement because there are never more than $n$ eigen-queries.  Thus Program~\ref{prog:expdesign}, $\appstrategy{\dsQ}{\W}$, has complexity $O(n^4)$, which is $O(n^4)$ times faster than solving $\optstrategy{\W}$.  We also find that the eigen-queries meet our expressiveness objective.   We will show this next by proving a bound on the approximation ratio.  In Sec. \ref{sec:eff} we propose techniques that exploit the fact that using subsets of the eigen-queries retain much of the expressiveness and increase efficiency.  And in Section \ref{sec:exp}, we show experimentally that weighted eigen-queries allow for near-optimal strategies, and also that the eigen-queries outperform other natural alternatives for the design set.

\subsection{The Eigen-Design Algorithm}

It remains to define the complete Eigen-Design algorithm, which is Program \ref{prog:eigen}:

\begin{algorithm}[h!]
\small
\begin{algorithmic}[1]
\REQUIRE Workload matrix $\W$.
\ENSURE Strategy matrix $\A$.
\STATE Compute the eigenvalue decomposition of $\W^T\W=\Q^T\D\Q$, where $\D=diag(\sigma_1, \ldots, \sigma_n)$ and set $\dsQ=\Q$.
\STATE Compute weights $\lambda_1,\ldots,\lambda_n$ by solving Program~\ref{prog:expdesign} for above $\dsQ$ and with $c_i=\sigma_i$, $i\in [1..n]$.
\revision{
\STATE Construct matrix $\M=\Q^T\LambdaB\Q$ where $\LambdaB$ is the diagonal matrix whose entries are $\lambda^2_1,\ldots,\lambda^2_n$.
\STATE Let $m_{11}, \ldots, m_{nn}$ be the diagonal entries of $\M$ and define $\M'$ to be the diagonal matrix whose entries are $\max_i\{m_{ii}-m_{11}\},\ldots, \max_i\{m_{ii}-m_{nn}\}$.
\STATE Compute the eigenvalue decomposition of $\M+\M'=\Q'^T\D'\Q'$, where $\D'=diag(\sigma'_1, \ldots, \sigma'_n)$ and return $\A=diag(\sqrt\sigma'_1, \ldots, \sqrt\sigma'_n)Q'$.}{
\STATE Construct matrix $\A'=\LambdaB\Q$ where $\LambdaB=diag(\lambda_1,\ldots,\lambda_n)$.
\STATE Let $m_{11}, \ldots, m_{nn}$ be the $L_2$ norm of columns of $\A'$ and define $\D'=diag(\max_i\{\sqrt{m^2_{ii}-m^2_{11}}\},\ldots, \max_i\{\sqrt{m^2_{ii}-m^2_{nn}}\})$.
\STATE Return $\A=\left[\begin{smallmatrix}\A'\\ \D'\end{smallmatrix}\right]$.
}
\end{algorithmic}
\caption{The Eigen-Design Algorithm}
\label{prog:eigen}
\end{algorithm}

The algorithm performs the decomposition of $\W^T\W$ to derive the design queries (Step 1), and solves $\appstrategy{\dsQ}{\W}$ using the eigen-queries as the design set (Step 2).  \revision{The matrix $\M$ that is constructed in Step 3 is a candidate matrix for $\A^T\A$.  But the output of the algorithm after Step 3 may result in a strategy with one or more columns whose norm is less than the sensitivity.  In this case, it is possible to add queries to the strategy without raising the sensitivity (Step 4), and these additional queries can only provide more information about the database, and hence reduce error.  Step 5 decomposes the revised candidate for $\A^T\A$ in order to derive the final strategy matrix.}{The matrix $\A'$ that is constructed in Step 3 is a candidate strategy but may have one or more columns whose norm is less than the sensitivity. In this case, it is possible to add queries, completing columns, without raising the sensitivity (Step 4 and 5).  These additional queries can only provide more information about the database, and hence reduce error.}


\subsection{Analysis of the Eigen-Design Algorithm}\label{sec:alg:analysis}

We now consider the accuracy and generality of the eigen-design algorithm, showing a bound on the worst-case approximation rate and that the accuracy of the algorithm is robust with respect to the representation of the input workload.

\subsubsection*{Approximation Rate}
To bound the approximation rate, we use an existing result showing a lower bound on the optimal error achievable for a workload using the $(\epsilon, \delta)$-matrix mechanism~\cite{Li11Measuring}.  The existence of this bound does not imply an algorithm for achieving it, but it is a useful tool for understanding theoretically and experimentally the quality of the strategies produced by $\appstrategy{}{\W}$ using the eigenvalues of $\W$. 
\begin{theorem}{\sc(Singular Value Bound~\cite{Li11Measuring})}\label{thm:svdb}
Given \\any $m\times n$ workload $\W$. Let $\sigma_1, \ldots, \sigma_n$ be the eigenvalues of matrix $\W^T\W$. The singular value bound of $\W$ is $\svdb(\W)\!\!=\!\!\frac{1}{n}(\sqrt{\sigma_1}+\ldots+\sqrt{\sigma_n})^2$,\! and bounds $\opterror{\W}$:
\revision{
\[\minerror(\W)\geq P(\epsilon, \delta)\svdb(\W).\]
}{
\[\opterror{\W}\geq \sqrt{P(\epsilon, \delta)\svdb(\W)}.\]}
\end{theorem}

Intuitively, let $\A_l$ be the strategy that is defined by weighting the eigen queries of $\W$ by $\sqrt{\sigma_1},\ldots,\sqrt{\sigma_n}$. The singular value bound comes from underestimating the sensitivity of $\A_l$ using $\sqrt{\tr(\A_l^T\A_l)/n}$. In practice, though the singular value bound may not be achieved since there is a gap between the sensitivity of $\A_l$ and $\sqrt{\tr(\A_l^T\A_l)/n}$, the idea of weighting the eigen queries can be combined with the experimental design method to find good strategies to $\W$.

Notice the strategy $\A_l$ is contained in the possible solutions of Program~\ref{prog:eigen}.  Thus the approximation ratio of Program~\ref{prog:eigen} can be estimated by using the approximation ratio of the singular value bound.
\begin{theorem}
Given workload $\W$, let $\sigma_1$ be the largest \\eigenvalue of $\W^T\W$, Program~\ref{prog:eigen} gives a strategy that approximates \revision{the minimal error of $\W$}{$\opterror{\W}$} with a ratio of \revision{$\sqrt{n\sigma_1/\svdb(\W)}$}{$(n\sigma_1/\svdb(\W))^{1/4}$}.
\end{theorem}
This theorem shows that the approximation ratio of applying Program~\ref{prog:eigen} to a workload $\W$ can be bounded by analyzing the eigenvalues of matrix $\W^T\W$.

In practice, \revision{the gap between $\svdb(\W)$ and $\apperror{}{\W}$}{the ratio between the error of the eigen strategies and the optimal error} is much smaller for a wide range of common workloads.  In the experiments in Sec. \ref{sec:exp}, \revision{the largest ratio to $\svdb(\W)$}{the largest ratio} is \revision{$1.6$}{at most $1.3$} and in a number of cases \revision{$\apperror{}{\W}$}{the ratio} is essentially equal to \revision{$\svdb(\W)$}{1}, modulo numerical imprecision.  

\subsubsection*{Representation Independence}

We say that the Eigen-Design algorithm is representation independent because its output is invariant for semantically equivalent workloads and error equivalent workloads.  Recall that the logical semantics of a workload matrix $\W$ depends on its cell conditions. \revision{For any $\W$, there are many different matrices which can represent a semantically equivalent workload if we consider an alternative definition of the cell conditions.  As a simple example, if we reorder the cell conditions of $\W$, then a new matrix with accordingly reordered columns will be semantically equivalent to the original.}{For any workload matrix $\W$, reordering its cell conditions leads to a new matrix $\W'$ with accordingly reordered columns. In this case, we say $\W$ and $\W'$ are \textsl{semantically-equivalent}.}

Naturally, we hope for a mechanism with equal error for any two semantically-equivalent representations of a workload.  Some prior approaches do not have this property.  For example, the wavelet and hierarchical strategies exploit the locality present in the canonical representation of range queries.  An alternative matrix representation of the range queries may result in significantly larger error.  The Eigen-Design algorithm does not suffer from this pitfall: 

\begin{proposition}[Semantic equivalence]
Let $\W_1$ \\and $\W_2$ be two semantically-equivalent workloads and suppose Prog.~\ref{prog:eigen} computes strategy $\A_1$ on workload $\W_1$ and $\A_2$ on workload $\W_2$. Then $\error{\A_1}{\W_1}=\error{\A_2}{\W_2}$.
\end{proposition}
\revision{\begin{proof} (Sketch) For any two semantically-equivalent workload matrices $\W_1$ and $\W_2$, there exist transformation matrices $\T_1$ and $\T_2$ such that $\W_1=\T_1\W_2\T_2$ where $\T_1$ performs row swaps and $\T_2$ performs a sequence of column swaps, column duplications, or duplicate column elimination.  Because $\T_1$ is actually an orthogonal matrix, $\W_2^T\W_2=(\T_1\W_2)^T(\T_1\W_2)$. In addition, the operations on $\T_2$ do not change the nonzero of eigenvalues of $\W_2^T\W$ and using $\dsQ\T_2$ instead of $\dsQ$ in Program~\ref{prog:expdesign} does not change the inequality constraint w.r.t. those $x_i$ that have non-zero eigenvalues. Therefore, Program~\ref{prog:expdesign} computes semantically-equivalent strategies $\A\T_2$ and $\A$ for $\W_1$ and $\W_2$, respectively, and the final step in Program~\ref{prog:eigen} will leave the strategies semantically-equivalent as well. Thus $\A_1=\A_2\T_2$ and $\error{\A_1}{\W_1}=\error{\A_2}{\W_2}$.
\end{proof}}{}

A related issue arises for two workloads that may be semantically different, but can be shown to have equivalent error.  Since $\W$ appears as $\W^T\W$ in the expression for \revision{total error}{error} of a workload, it follows that, for any orthogonal matrix $\Q$, workload $\Q\W$ has error equal to $\W$ under any strategy.  And in particular, any two such workloads have equal minimum error.  The Eigen-Design algorithm always finds the same strategies for any two error-equivalent workloads:

\begin{proposition}[Error equivalence]
Let $\W_1$ and \\$\W_2$ be two error-equivalent workloads (i.e. $\W_1=\Q\W_2$ for some orthogonal $\Q$) and suppose Program~\ref{prog:eigen} computes strategy $\A_1$ on workload $\W_1$ and $\A_2$ on workload $\W_2$. Then $\error{\A_1}{\W_1}=\error{\A_2}{\W_2}$
\end{proposition}

This result follows from the fact that the input to Program~\ref{prog:expdesign} uses the eigenvectors of $\W^T\W$, and therefore operates identically on equivalent workloads.  

\revision{}{\subsubsection*{Optimizing for Relative Error}
The discussion above is about workload error, an {\em absolute} measure of error.  Our adaptive approach can also be used to find strategies offering low {\em relative} error.  However, these are two fundamentally different optimization objectives and a single strategy matrix will not, in general, satisfy both.

One major difference between computing absolute error and relative error is the impact of the $L_2$ norm of a query vector. According to Prop.~\ref{def:m-mech} and Def.~\ref{def:errors}, the query error of $\w$ under strategy $\A$ is proportional to the $L_2$ norm of $\w$. Therefore a scaled query $k\w$ has $k$ times larger query error compared with $\w$, and thus a query with higher $L_2$ norm contributes more to workload error. But because the relative error does not change with the $L_2$ norm of the query, using strategies optimized for workload error will not lead to optimal relative error.

Because the matrix mechanism is a data-independent mechanism, it is not possible to optimize for relative error directly.  If the distribution of the target dataset were known, we could scale each query by its weighted $L_2$ norm, where the weight on each cell is proportional to the inverse of its probability.  This scaling will optimize towards relative error by neutralizing the fact that the designed strategies are biased towards high norm queries.  Since the underlying distribution is typically unknown, we introduce a heuristic scaling, prior to applying the Eigen-Design algorithm, in which each query is normalized to make its $L_2$ norm $1$. This is equivalent to assuming a uniform distribution over the cells.  In Sec~\ref{sec:exp}, we show that, for two real datasets, this approach results in significantly lower relative error than competing techniques.}

\subsection{Application to the $\epsilon$-Matrix Mechanism} \label{sec:sub:l1}

There are a number of challenges to applying the optimally weighted design approach under $\epsilon$-differential privacy.  Recall, once again, the formula for \revision{total error}{workload error}  from Prop. \ref{prop:totalerror}: \revision{$|| \A ||_2^2 \;\tr (\W^T\W(\A^T\A)^{-1})$\\}{\!\!$|| \A ||_2\sqrt{\tr (\W^T\W(\A^T\A)^{-1})}$}. To move to $\epsilon$-differential privacy, only the sensitivity term changes, from $L_2$ to $L_1$: \revision{$|| \A ||_1^2 \;\tr (\W^T\W(\A^T\A)^{-1})$\\}{$|| \A ||_1 \sqrt{\tr (\W^T\W(\A^T\A)^{-1})}$}.  In the former case, the sensitivity term $|| \A ||_2$ is uniquely determined by $\A^T\A$.  But in the latter case, computing a near-optimal $\A^T\A$ is not enough, because $|| \A ||_1$ remains undetermined and is itself hard to optimize. As a result, it is more challenging (although still possible) to represent the optimal query weighting as a convex optimization problem.  We omit its formal encoding, but note that the resulting problem is also less efficient because we can no longer rely on second order cone programming.

\revision{Even though we can solve the problem using convex optimization, it is not clear that the eigen-queries will serve as a good design set since they characterize only the properties of $\W^T\W$ but do not account for the $L_1$ sensitivity.}{Furthermore, there does not seem to be a universally good design set: the eigen-queries do not outperform other bases, in general, because they characterize only the properties of $\W^T\W$ but do not account for the $L_1$ sensitivity.  
We can nevertheless still use our algorithm to improve existing strategies.  For example, using the Wavelet basis in the algorithm can improve its performance on all range and random range queries by a factor of $1.1$ and $1.5$, respectively; using the Fourier basis can improve its performance on low order marginals by a factor of $1.6$.} 

Lastly, \revision{there is no}{we do not know of an} analogue of Thm \ref{thm:svdb} providing a guaranteed error bound for the $\epsilon$-Matrix Mechanism to verify the quality of the output.  

These challenges motivate our choice to focus on $(\epsilon, \delta)$-differential privacy.  While the two privacy guarantees are strictly-speaking incomparable, for conservative settings of $\delta$, a user may be indifferent between the two.  It is then possible to show that the asymptotic error rates for many workloads are roughly comparable between the two models. 

\section{\revision{Performance}{Complexity And} Optimizations} \label{sec:eff}

We focus next on methods to further reduce the complexity of approximate strategy selection.  \revision{We first show that for workloads with low rank, the strategy selection algorithm can be solved more efficiently, }{We first analyze the complexity of the strategy selection algorithm and show that it can be solved more efficiently for low rank workloads, }with no impact on the quality of the solution.  Then we propose two approaches which can significantly speed up strategy selection by reducing the size of the input to Program~\ref{prog:eigen}.  Intuitively, both approaches perform strategy selection over a summary of the workload that is constructed from its most significant eigenvectors, potentially sacrificing fidelity of the solution.  We evaluate the latter two techniques in Sec \ref{sec:exp:tradeoff}.

\subsection{\revision{Workloads without full rank}{Complexity Analysis}}
The rank of workload matrix $\W$, denoted by $\rank(\W)$, is the size of the largest linearly-independent subset of the rows (or, equivalently, columns).  When $\rank(\W)$ is its maximum value, $n$, we say that $\W$ has full rank, which implies that accurate answers to the workload queries in $\W$ uniquely determine every cell count in $\x$.  \revision{}{The complexity of the strategy selection algorithm can be broken into three parts: computing the eigenvectors and eigenvalues of matrix $\W^T\W$, solving the optimization problem, and constructing the strategy. If an eigenvalue is equal to zero, the eigenvalue and its corresponding eigenvectors are not actually involved the optimization and strategy construction, so they can be omitted in practice. Since the number of nonzero eigenvalues of $\W^T\W$ is equal to $\rank(\W)$, the complexity of Programs~\ref{prog:eigen} is $O(nm\,\rank(\W)+n\,\rank(\W)^3)$.

The complexity analysis above indicates that its efficiency can be significantly improved when $\rank(\W)\ll n$.}\revision{
Workloads often have lower rank, especially when they are associated with multi-dimensional data sets.  In this case, intuitively, some groups of cell counts always appear together and could be treated as a single variable. For example, the workload in Fig. \ref{tbl:one}(b) has 8 columns but rank $4$, and is therefore not full rank.  In addition, workloads of low-order marginals are not full rank.

We can take advantage of the low rank of a workload to improve the efficiency of our algorithm.  Observing Programs~\ref{prog:expdesign} and \ref{prog:eigen}, if an eigenvalue of matrix $\W^T\W$ is 0, its corresponding $c_i$ in Program~\ref{prog:expdesign} will be $0$. When $c_i=0$, we can let $v_i$ be arbitrarily large and $u_i$ be arbitrarily small so that the optimal solution to Program~\ref{prog:expdesign} must have $u_i=0$ for any $c_i=0$. Therefore, we can directly remove this $u_i$ and its corresponding $v_i$ from the convex optimization problem to reduce the number of variables by 2 and the number of constraints by 1.  A non-full rank workload $\W$ has exactly $\rank(\W)$ non-zero eigenvalues, and the complexity of solving Program~\ref{prog:eigen} can be reduced to $O(n\,\rank(\W)^3)$.

The efficiency of the algorithm can be significantly improved when $\rank(\W)\ll n$.}{}  For example, the rank of low order marginal workloads can be bounded by the number of queries in the workload. Suppose a low-order marginal workload is defined on a $k$-dimensional space of cell conditions, each of which has size $d$.  If the workload only contains one-way marginals, the complexity of solving Program~\ref{prog:eigen} over this workload is bounded by $O(k^3d^{3+k})$.  If the workload consists of one and two-way marginals the complexity is $O(k^6d^{k+6})$. Both of these bounds are much smaller than $O(d^{4k})$\revision{, the complexity of running Program~\ref{prog:eigen} directly}{}.

\subsection{Workload Reduction Approaches}

Next we propose two approaches which allow us to reduce the number of variables in the optimization problem. Both are inspired by principal component analysis (PCA), in which a matrix is characterized by the so-called principal eigenvectors, which are the eigenvectors associated with the largest eigenvalues.  


In our case, recall that we cannot ignore the non-principal eigenvectors since the rank of the strategy matrix $\A$ cannot be lower than the workload matrix $\W$.  Instead, we either compute separately the weights for the principal and remaining eigenvectors, or we choose the same weights for all the remaining eigenvectors.

\subsubsection*{Eigen-Query Separation}

In {\em eigen-query separation}, we partition the eigen-queries into groups of a specified size according to their corresponding eigenvalues.  Treating one group at a time, Program~\ref{prog:expdesign} is executed to determine the optimal weights just for the eigenvectors of that group.  After the individual group optimizations are finished, another optimization can be used to calculate the best factor to be applied to all queries in each group.  If the group size is large, all of the principal eigenvectors may be contained in one group, in which case the most important weights will be computed precisely. 

The complexity of eigen-query separation depends on the group division. Notice that during the optimization of each group, the convex optimization problem is equivalent to setting all eigenvalues of excluded eigenvectors to zero.  Analogous to the discussion of low rank workloads, letting the size of group be $n_g$, the complexity of solving the optimization problem over each group is $O(nn_g^3)$. Similarly, the time complexity to combine all the groups is $O(n(n/n_g)^3)$, and therefore $O(n^2n_g^3+n(n/n_g)^3)$ in total. Asymptotically, the complexity of eigen-query separation is minimized when $n_g=O(n^{1/3})$.  Then the complexity of the entire process is $O(n^{3})$, the same as the cost of standard matrix multiplication.

\subsubsection*{Principal Vectors Optimization}
In the {\em principal vector optimization} we use a subset of the $k$ most important eigenvectors as the design set, computing the optimal weights as usual.  Instead of ignoring the less important eigenvectors (as is typical in PCA) we simply use a single common weight for each of the excluded vectors that have non-zero eigenvalues.  The number of variables in the convex optimization is reduced to $k+1$ so that the time complexity is reduced to $O(nk^3)$.  Experimentally we find that good results are possible with as little as $10\%$ of the eigenvectors.

\vspace{1ex}
In Sec.~\ref{sec:exp:tradeoff} we show that both of the above approaches can improve execution time by two orders of magnitude with modest impact on solution quality.  Extending our theoretical bound on the approximation rate to these approaches is an interesting direction for future work.

\section{Experimental Evaluation} \label{sec:exp}

\subfigcapskip=-7pt 

The empirical evaluation of our mechanism has three objectives: $(i.)$ to measure \revision{}{solution quality of the Eigen-Design algorithm using both absolute and relative error}; $(ii.)$ to measure the trade-off between speed-up and solution quality of our two performance optimizations; and $(iii.)$ to measure the effectiveness of using the eigen-queries as the design set.  Experimental conclusions are presented in Sec. \ref{sec:sub:con}.

%
%
%
\begin{figure*}[th]
\centering
\subfigure[\small Absolute errors on range queries]{
	\includegraphics[height=100pt]{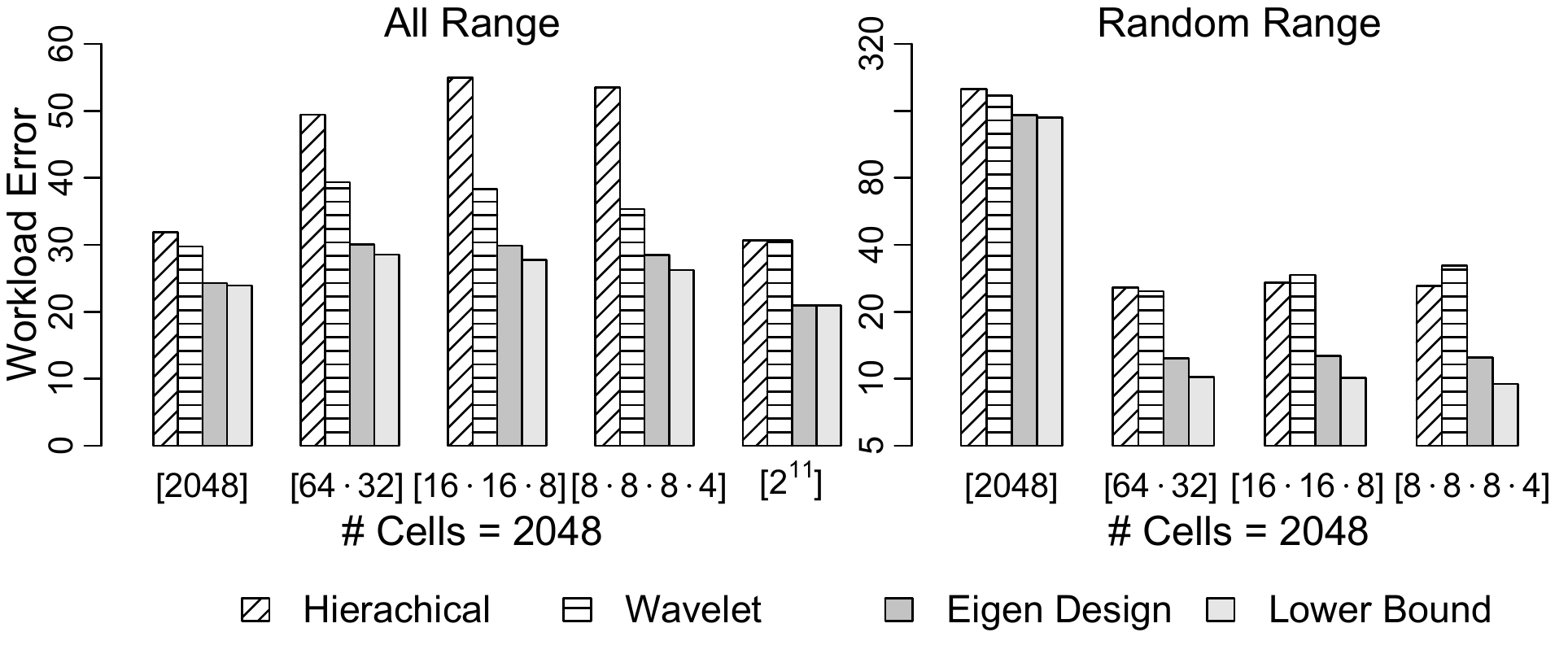}
	\label{fig:range}}
\subfigure[\small \revision{}{Relative errors on range queries}]{
	\includegraphics[height=100pt]{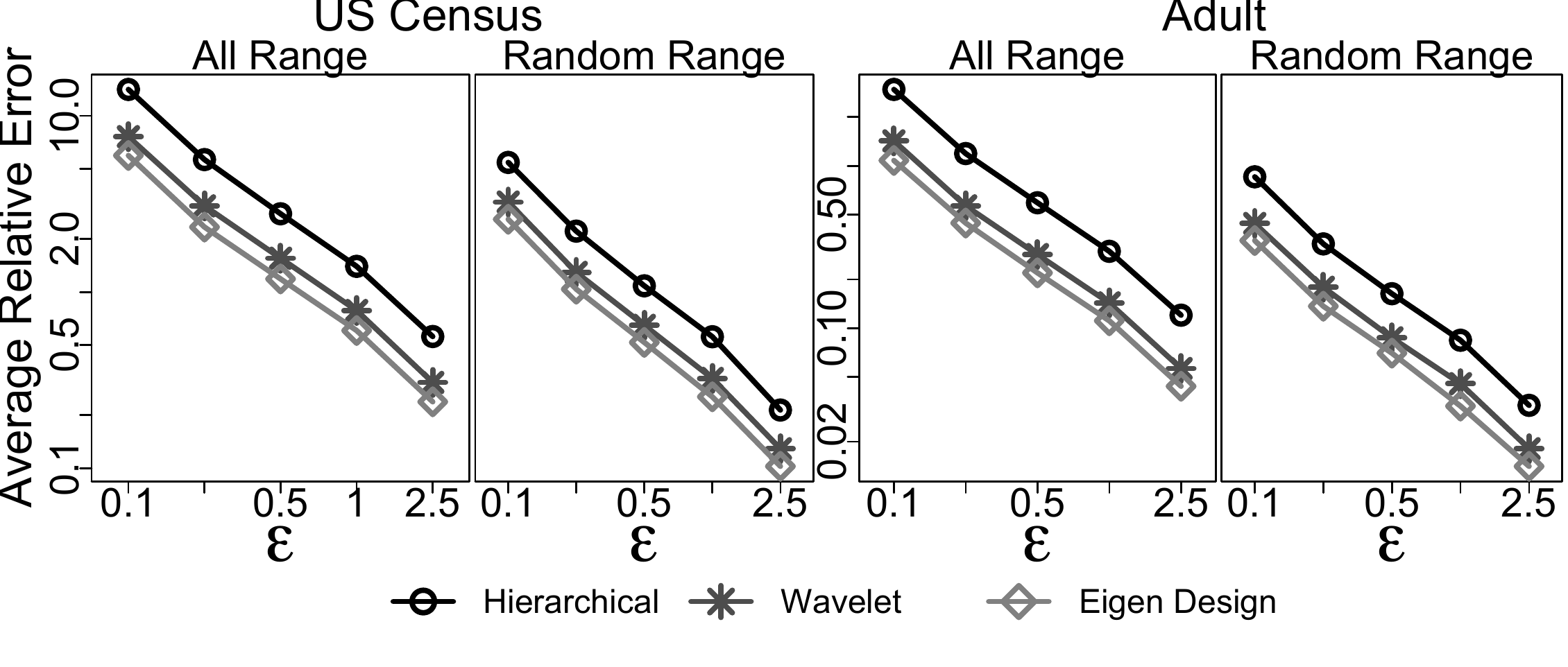}
	\label{fig:realrange}}\\
	\vspace*{3pt}
\subfigure[\small Absolute errors on marginal queries]{
	\includegraphics[height=96pt]{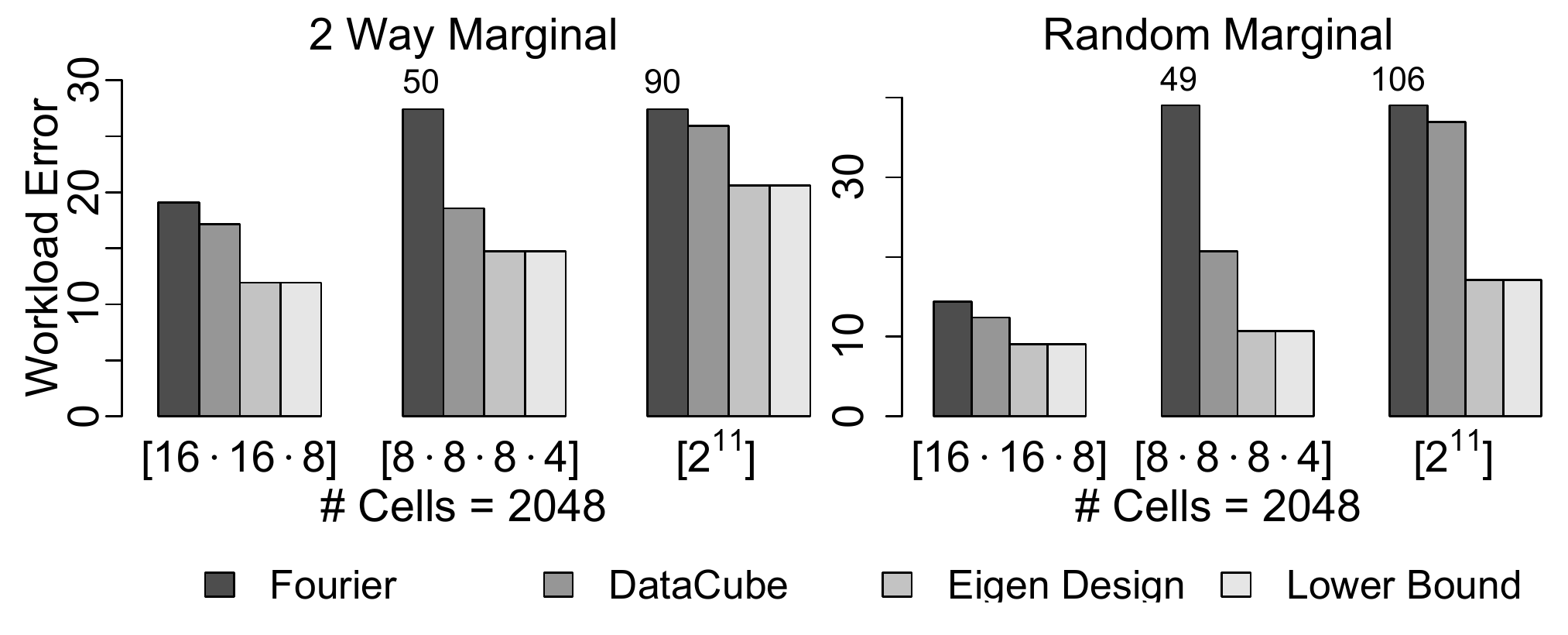}
	\label{fig:marginal}	}
\subfigure[\small \revision{}{Relative errors on marginal queries}]{
	\includegraphics[height=100pt]{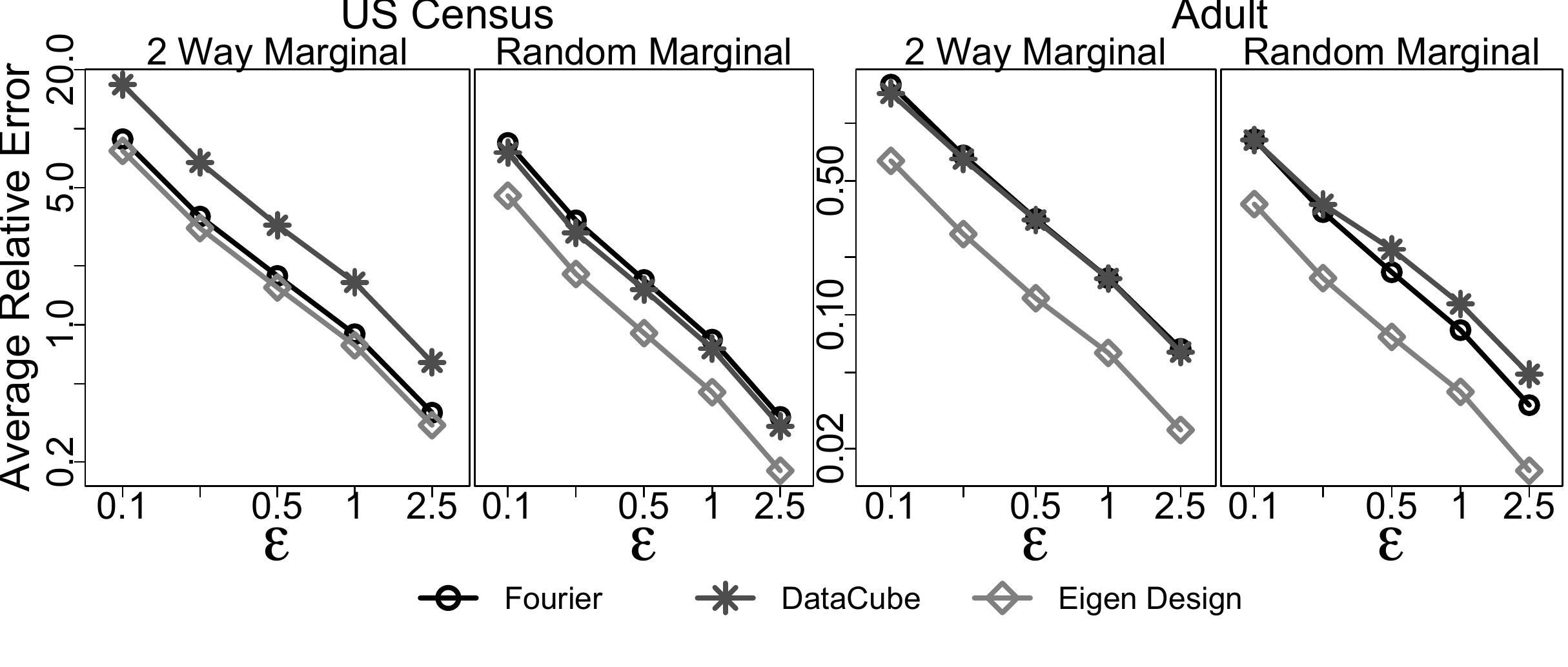}
	\label{fig:realmarginal}}
	\vspace*{-4pt}
\caption{\revision{}{Absolute and relative error for the Eigen-Design algorithm and competitors, for range and marginal workloads, on 2048 cells.  {\sf ``Lower Bound''} is a bound on the best possible error achievable by any strategy.}}\label{fig:regular}
	\vspace*{-8pt}
\end{figure*}

\subsubsection*{Experimental Setup}

Recall that \revision{root mean squared error}{workload error is an absolute error measure based on root mean square error}.  \revision{}{Workload} error can be analytically computed using Prop. \ref{prop:totalerror}, and this is precisely the error that will be witnessed when running repeated trials and computing the mean deviation.  Further, workload error is independent of the true counts in data vector $\x$. That is, it is independent of the input data.  These facts hold for all instances of the matrix mechanism, and therefore for each of the competing techniques we consider below.   Therefore, \revision{}{when evaluating this absolute error measure}, we do not perform repeated trials with samples of random noise nor do we use any datasets.  In addition, all measures of workload error include the same factor $P(\epsilon, \delta)$, so that changing the privacy parameters impacts each method \revision{equally}{with the same factor}, leaving the ratio of their error the same.  Consequently, \revision{}{for workload error}, we simply \revision{assume $P(\epsilon, \delta)=1$}{fix $\epsilon=0.5$ and $\delta=0.0001$}. 

For \revision{}{workload} error, all error measurements are purely a function of the workload, reflecting the hardness of simultaneously answering a set of queries under differential privacy.  In addition, these error rates can be compared directly with the lower bound as Theorem~\ref{thm:svdb}, reflecting a bound on the approximation rate. (This lower bound is not known to be achievable for all workloads, but nevertheless informs the quality of the eigen-strategy and its competitors.)  

\revision{We do not present relative error measures, which depend on the properties of an input dataset, however we have observed relationships between the techniques that are similar to those presented below.}{We also evaluate the relative error rates achievable using our algorithm by computing the strategy that minimizes absolute error on a scaled workload, as described in Sec.~\ref{sec:alg:analysis}.  Of course, the relative error rates reported in experiments are always for the original input workload. 
In these experiments we vary the value of $\epsilon$, for a fixed $\delta=0.0001$, and consider two real datasets.  The first dataset is the US individual census data in the past five years\footnote{Integrated Public Use Microdata Series: usa.ipums.org}, which are aggregated on age, occupation and income. The second is the Adult dataset\footnote{UCI Machine Learning Repository: archive.ics.uci.edu/ml/}, in which tuples are weight-aggregated on age, work, education and income. The size and dimensions of the datasets are: \vspace{-1ex}
\begin{table}[h]
\centering
\revision{}{
\begin{tabular}{|c|c|c|}
\hline Dataset & Dimension & \# Tuples \\
\hline US Census & $8\times 16\times 16$  & 15M \\
\hline Adult & $8\times 8\times 16\times 2$ & 33K \\
\hline
\end{tabular}}
\vspace*{-1ex}
\caption{The size and dimensions of the datasets}
\end{table}
\vspace*{-1ex}
}





All experiments are executed on a quad-core $3.16$GHz Intel CPU with $8$ GB memory.  Our Python implementation extends publicly-available code for the matrix mechanism \cite{mmech-code} and also uses the {\tt dsdp} solver \cite{dsdp5} in the {\tt cvxopt} \cite{cvxopt} package.

\subsubsection*{Competing Approaches}
We compare the Eigen-Design strategy with the following four alternatives. Although originally proposed in the context of $\epsilon$-differential privacy, each is easily adapted to $(\epsilon, \delta)$-differential privacy and the shift  generally improves the relationship to the optimal error rate (with the exception of the Fourier strategy, noted below).

\begin{description} \itemsep 0in

\item[]\textbf{Fourier} is designed for workloads consisting of all $k$-way marginals, for given $k$~\cite{barak2007privacy}. The strategy transforms the cell counts with the Fourier transformation and computes the marginals from the Fourier parameters. When the workload is not full rank, the unnecessary queries of the Fourier basis are removed from the strategy to reduce sensitivity.  The effectiveness of the Fourier strategy is somewhat reduced under $(\epsilon, \delta)$-differential privacy because dropping unnecessary queries results in a smaller sensitivity reduction using $L_2$.

\item[]\textbf{DataCube} is an adaptive method that supports marginal workloads~\cite{Ding:2011fk}. We implemented the BMAX algorithm, which chooses a subset of input marginals so as to minimize the maximum error when answering the input workload.  To adapt the algorithm to $(\epsilon, \delta)$-differential privacy,  sensitivity is measured under $L_2$ instead of $L_1$.

\item[]\textbf{Wavelet} supports multi-dimensional range workloads by applying the Haar wavelet transformation to each dimension~\cite{xiao2010differential}. When using $\epsilon$-differential privacy, Xiao et al. also introduced a hybrid algorithm that uses the identity strategy on dimensions with small size.  This optimization is unnecessary under $(\epsilon,\delta)$-differential privacy: the hybrid algorithm does not lead to smaller error when sensitivity is measured under $L_2$.

\item[]\textbf{Hierarchical} aims to answer workloads of range queries using a binary tree structure of queries: the first query is the sum of all cells and the rest of the queries recursively divide the first query into parts~\cite{Hay:2010Boosting-the-Accuracy}. We test binary hierarchical strategies (although higher orders are possible).  The strategy in \cite{Hay:2010Boosting-the-Accuracy} supports one dimensional range workloads, but is adapted to multiple dimensions in a manner analogous to Wavelet \cite{xiao2010differential}. 
\end{description}

We do not compare with the error of the standard Gaussian mechanism, which, for the workloads considered, is far worse than all alternatives.  Prior works \cite{Hay:2010Boosting-the-Accuracy,xiao2010differential,Ding:2011fk} compared the error rates of their approaches with the identity strategy.  We omit this explicit comparison, since the identity is always within the space of possible strategies the Eigen-Design could choose, but is not competitive.


\subsection{Error of the Eigen-Design Algorithm}
We now measure the improvement in \revision{}{absolute and relative error} offered by the Eigen-Design algorithm along with its approximation to optimal \revision{}{absolute} error.  Below we refer to the strategy produced by the Eigen-Design algorithm, for a given workload, as the {\em eigen-strategy}.  We consider three classes of workloads, \revision{beginning with structured workloads, then workloads of randomly sampled queries,}{beginning with workloads of range queries, then workloads of marginals}, and then some alternative workloads designed to test the adaptivity of the mechanism.  

\revision{
\paragraph*{Structured workloads} 
\textbf{Figs.~\ref{fig:regular}(a)-(c)} contain experiments on workloads of all range queries, all marginal queries, and all two-way marginal queries.  For all range queries, the eigen-stratgy is compared with Hierarchical and Wavelet. For workloads of marginals, the eigen-strategy is compared with DataCube and Fourier.  

The results show that the eigen-design strategies reduce error by at least a factor of 2 compared to the best competing strategies in most cases.  In addition, the error of eigen-design strategy is very close to (in \ref{fig:allrange}) or reaches (in \ref{fig:allmarginal} and \ref{fig:lowmarginal}) the lower bound of error, showing that it is not possible to improve much on the eigen-strategy.  The error of the Fourier strategy is not fully shown in the high dimensional cases of \ref{fig:lowmarginal} since it is worse than all other strategies by more than an order of magnitude.

\paragraph*{Randomized workloads}

Next we consider randomly selected low-order marginal queries and randomly selected three dimensional range queries.  For workloads of marginal queries, in \textbf{Fig.~\ref{fig:rmarginal}}, we randomly sample several attributes or pairs of attributes and use all marginal queries over the selected attributes. The result over the random sampled marginal queries is quite similar to the two way marginal case, but the gaps between the eigen-design strategy and the competitors are even larger: the errors are reduced by a factor of 5 compared with the DataCube strategy and the Fourier is about 6 times worse than DataCube.

For workloads of random range queries, we follow the two-step sampling method in \cite{xiao2010differential}: the first step chooses the dimensions of the sampling query uniformly at random, and the second step picks one query from all queries over the selected dimensions uniformly at random.  Those randomly generated queries are partitioned into five equal groups according to their coverage percentage of the domain, and each forms a workload.  

As shown in \textbf{Fig. \ref{fig:privlet}}, we compare the eigen-design strategy, computed for each individual workload, to the Wavelet and the Hierarchical strategy.  We also compute the eigen-design strategy for the union of the five workloads to form a ``universal'' eigen-design strategy. The results indicate that the universal eigen-design strategy introduces error almost evenly to each group and consistently reduces the error by up to a factor of 2.5 compared to competitor strategies.  The eigen-design strategies designed for each group are even better, and clearly show the benefit of an adaptive algorithm.  Error is reduced by a factor of 5 compared to the Hierarchical strategy, and outperforms the Wavelet strategy by about an order of magnitude.}
{
\paragraph*{Workloads of Range Queries} 
\textbf{Figs.~\ref{fig:regular}(a),(b)} contain experiments on workloads of all range queries and random range queries. The random ranges are sampled with the two-step sampling method in \cite{xiao2010differential}. Here the eigen-strategies are compared with Hierarchical and Wavelet strategy. The figures are in log scale, except Fig.~\ref{fig:regular}(a) on all range queries. The results show that the eigen-design strategies reduce error by a factor of 1.2 to 2.1 in workload error and 1.3 to 1.5 in relative error compared to the best competing strategies. In addition, for workload error, the eigen-design strategy is within a factor of 1.3 to the lower bound.

\paragraph*{Workloads of Marginals}
\textbf{Figs.~\ref{fig:regular}(c),(d)} contain experiments on workloads of 2-way marginal queries and random marginal queries, in which the random marginals are sampled with the sampling method in \cite{Ding:2011fk}. Here the eigen-strategies are compared with Fourier and DataCube.  The figures are in linear scale for workload error and log scale for relative error. The results show that the eigen-design strategies reduce error by a factor of 1.3 to 2.2 compared to the best competing strategies in workload error, and by a factor of 1.1 to 2.7 in relative error. In addition, the error of eigen-design strategies match the lower bound of workload error, indicating that our algorithm found an optimal strategy with respect to workload error.}
\begin{table}[h]
\centering
\vspace*{-5pt}
\revision{}{
\small
\begin{tabular}{|p{40pt}||p{33pt}|c|p{17pt}|c|}
\hline \multirow{2}{46pt}{Workload}& \multicolumn{3}{c|}{Error Ratio}& \multirow{2}{35pt}{Best/Worst Competitor}\\
\cline{2-4} & \hspace*{-1pt}Err Type & Best/Worst  & \hspace*{-3.5pt}Bound & \\
\hline  \multirow{2}{46pt}{\hspace*{-3pt}1D Range\newline\hspace*{-3pt}(Permuted)}& workload & 9.62/13.16 &0.99 &Wav./Hier.\\
\cline{2-5} & relative & 1.51/2.43  &- & Wav./Hier.\\
\hline  \multirow{2}{47pt}{\hspace*{-3pt}1Way Range\newline\hspace*{-3pt}Marginal}& workload & 1.30/7.69 & 0.98 & D.Cube/Four.\\
\cline{2-5} & relative & 1.36/4.93 &- & D.Cube/Four.\\
\hline  \multirow{2}{47pt}{\hspace*{-3pt}2Way Range\newline\hspace*{-3pt}Marginal}& workload & 1.63/3.23 & 0.95 & Hier./Four.\\
\cline{2-5} & relative &  1.81/2.38 &- & Wav./D.Cube \\
\hline  \multirow{2}{46pt}{\hspace*{-3pt}1D CDF}& workload & 1.01/1.01 &0.80 & Wav./Hier.\\
\cline{2-5} & relative & 0.46/0.54 &- & Wav./Hier.\\
\hline  \multirow{2}{46pt}{\hspace*{-3pt}Predicate}& workload & 1.39/1.94 &1.00 & Wav./Four. \\
\cline{2-5} & relative & 1.42/3.55 &- & Four./Hier.\\
\hline
\end{tabular}}
	\vspace*{-8pt}
\caption{\revision{}{The factor of error reduced for the Eigen-Design algorithm w.r.t. the best/worst competitors strategies and the theoretical bound, for alternative workloads, on $2048$ cells.}}\label{fig:mix}
	\vspace*{-12pt}
\end{table}

\begin{figure*}[t!]
\subfigure[\small Approximation approches over all 1D ranges]{
	\includegraphics[height=90pt]{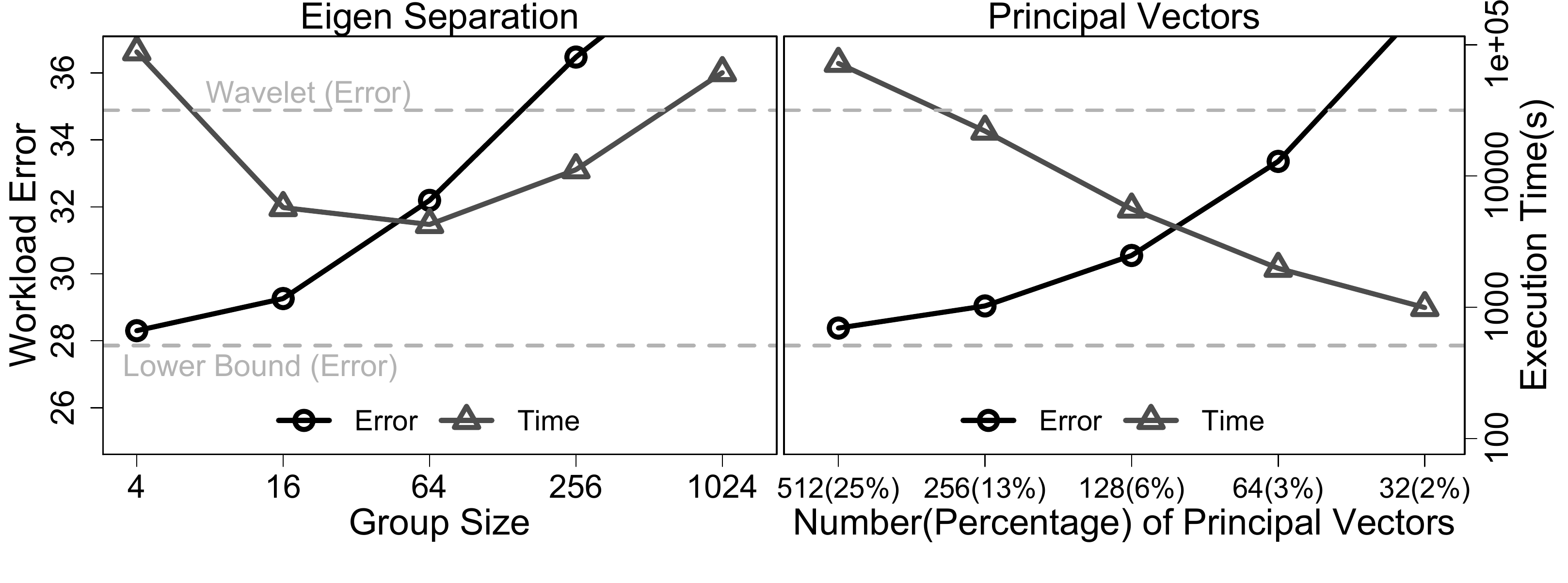}
	}
	\quad
\subfigure[\small Approximation approches over all 2D marginals]{
	\includegraphics[height=90pt]{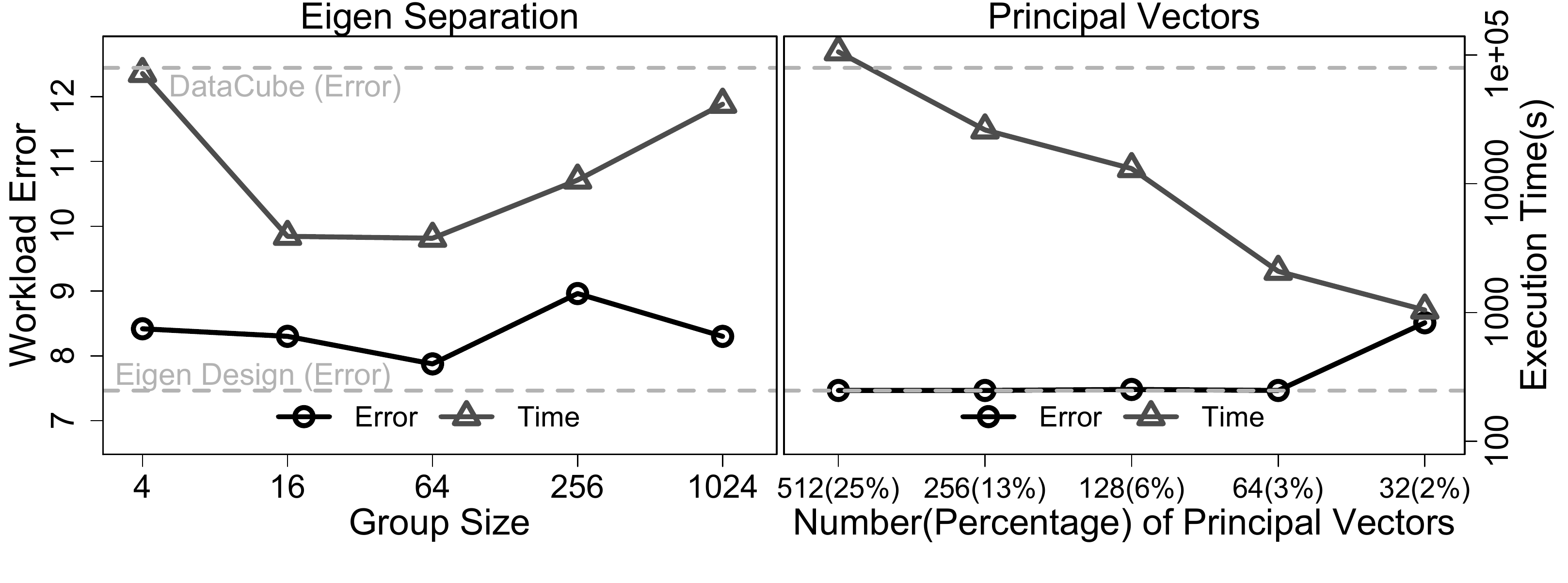}
	}
	\vspace*{-11pt}
\caption{\revision{}{Quality and efficiency of approximation methods on 8192 cell conditions}}\label{fig:effvsacc}
\vspace*{-12pt}
\end{figure*}

\paragraph*{Alternative Workloads}
To demonstrate that our mechanism is adaptive over variety of workloads, we also include other workloads that have not been studied in prior work. First we show that our mechanism adapts to semantically equivalent workloads, in which we repeat \revision{the experiments of a structured workload}{the experiment on range Workload} but randomly permute the order of cell conditions. The justification for this experiment comes from the fact that the user may wish to answer queries in which the order of the cell conditions is not obvious, such as predicate queries over categorial attributes.  \revision{Since the permutation destroys the structure of the marginals, the DataCube method can not be applied. In the experiments on permuted marginal queries, we use Wavelet instead. }{}

In addition, we run experiments on three other workloads: the range marginals workload, the cumulative distribution (CDF) workload, and uniformly sampled predicate queries. The range marginals workload is 
important because most data analyses using marginals do not simply use individual counts, but also aggregate counts.  If this is the case, simply computing the marginals workload privately is the wrong approach because error accumulates for aggregations. Last, the CDF workload is a highly-skewed set of one-dimensional range queries where the sensitivity in the first cell is $n$, decreasing linearly to 1 for the last cell.

We summarize the experimental results on alternative workloads in \textbf{Table~\ref{fig:mix}}. \revision{Since the error on different workloads differs greatly, we represent them as multiples of the lower bound. For the permuted workloads, the quality of the Hierarchical and the Wavelet strategy is greatly reduced, whose factor of error to the lower bound (cut off in the figure) is actually larger than \revision{90}{9} on all one-dimensional ranges. By contrast, the Fourier and Eigen-Design strategies retain the same performance as the original workload. }{For relative errors, due to space constraints, we only present results on US census data with $\epsilon=0.5$ and $\delta=0.0001$. We present, for each workload, the factor of error reduction achieved by our algorithm compared to the best and worst competing approach, whose name is shown in the last column of the table.  (Datacube is only considered for range marginals and Fourier is not considered on permuted range and CDF.)  In addition, for workload error, we also include the ratio to the error lower bound.}

\revision{In the comparison of the range marginal queries, we only include two of the best competing strategies: the Hierarchical and the DataCube strategy, whose errors are still as much as1.7 times the error of eigen-strategies.  For the CDF workload, the eigen-strategy is only slightly better than the Hierarchical and Wavelet workloads.  We suspect this is one case where the eigen-queries do not form a good design set due to the highly skewed nature of the workload.  Lastly, \revision{}{for predicate workload} the Eigen-Design strategy reduced the error of Fourier and Wavelet method by a factor of \revision{}{1.4 and 2}, respectively and introduces error that is almost identical to the lower bound.}{
The results show that the eigen-strategy can improve absolute error by as much as 13 times (on permuted range queries) and relative error as much as 5 times (on one-way range marginals).  The workload error of competing strategies is heavily impacted by the permutation but the relative errors are not as bad since queries of individual cells and small ranges dominate the workload, which do not change too much under permutation.  On all workloads but one, the eigen-strategy beats every competitor by at least a factor of 1.3, and is very close to---or achieves---the theoretical error lower bound.  The only exception is the CDF workload, in which the eigen-strategy is only a bit better than the competitor for workload error and worse (than Hierarchical and Wavelet) for relative error.  Overall, the results for workload and relative error are largely similar for range marginals and the predicate workload. 

}

\subsection{Performance Optimizations} \label{sec:exp:tradeoff}

\textbf{Fig.~\ref{fig:effvsacc}} illustrates the trade-off between computational \\speed-up and solution quality for the {\em eigen-separation} and {\em principal vector} performance optimizations described in Section~\ref{sec:eff}.  \revision{}{We only present results with workload errors here (the results with relative error are similar or even better). }Error and computation time are plotted together using two y-axes: the left axis measures \revision{average per query error}{workload error} and the right axis measures execution time in seconds.  \revision{The baseline for time measurements is the time for the standard Eigen-Design algorithm.  The baselines for error are the error of the standard Eigen-Design and the best competing technique.}{The baselines for error are the lower bound and the best competing technique.}

\revision{We find that both methods can reduce the running time by two orders of magnitude while increasing error by less than $20\%$ over the standard eigen-design solution.}{The running time of using the standard Eigen-Design algorithm can be estimated from the running time of the principal vector method, which is more than an order of magnitude slower than the principal vector method with $25\%$ of the eigenvectors. Comparing with this estimated time, both methods can reduce the running time by two orders of magnitude while the error they introduced is less than $12\%$ over the lower bound.} For the eigen-separation method, the computation in each group takes more time with larger group sizes while the computation of merging groups takes more time with smaller group sizes.  Theoretically, the best choice for group size of the eigen-separation method is $n^{1/3}$, which is closest to $16$ in this case.  Using eigen-query separation with a group size of 16, \revision{strategy selection is $300$ times faster, while the strategy found has roughly $12\%$ greater error.}{the error is $5\%$ higher on all range queries and  $11\%$ higher on all marginal queries.} Using the principal vectors optimization with $6\%$ of the eigenvectors, \revision{makes strategy selection $400$ times faster, while error is $13\%$ higher on all range queries and same as the optimal on all marginal queries.}{the error is $10\%$ higher on all range queries and the same as the optimal on all marginal queries.}

According to the results, the eigen-separation performs better on range queries while the principal vectors method is better on marginals.  In either case, the performance improvements still produce results that are significantly better than competing techniques.

\subsection{The Choice of Design Queries}

\begin{figure}[ht]
\centering
\includegraphics[height=80pt]{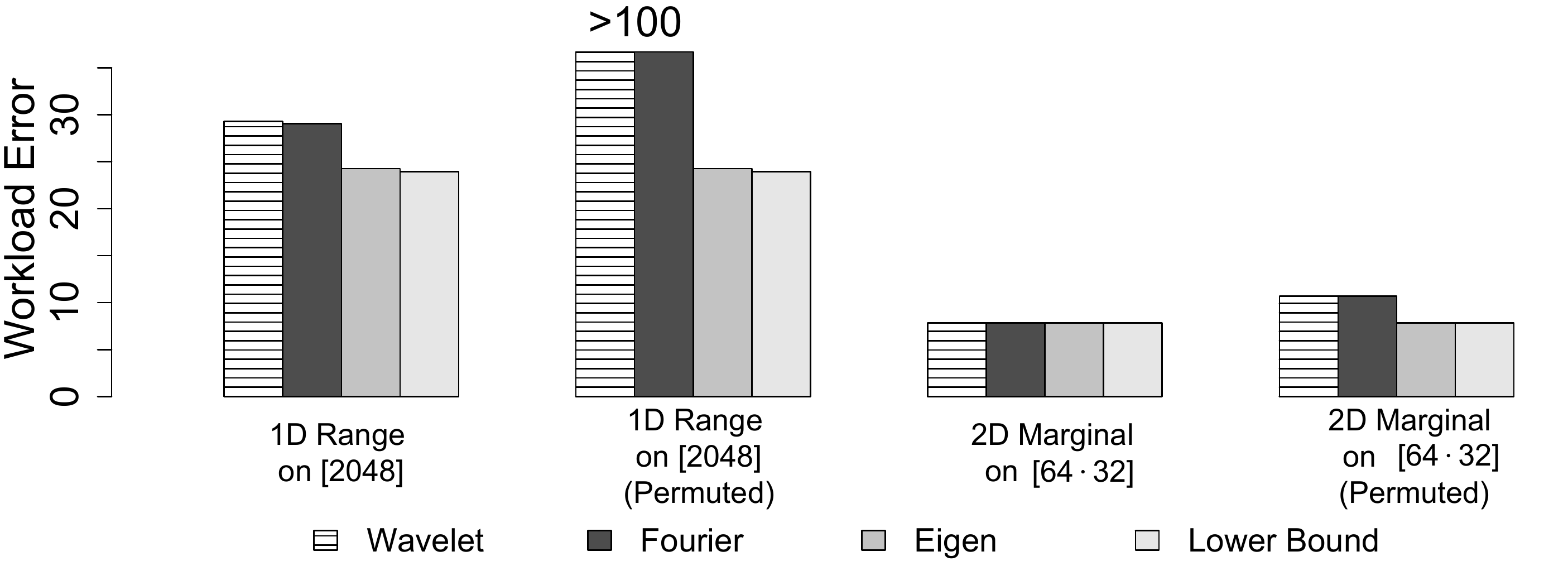}
\vspace*{-10pt}
\caption{Comparison of design queries}
\label{fig:basis}
\end{figure}

To evaluate our claim from Section~\ref{sec:alg:choose} that the eigen-queries are an effective choice for the design queries we compare strategies computed by Program~\ref{prog:expdesign} using the eigen-queries, the Wavelet matrix and Fourier matrix as the design queries. Since using the eigen-queries introduces the same error to semantically equivalent workloads, we also empirically verify this property on other sets of designed queries. \textbf{Fig.~\ref{fig:basis}} shows the results of those comparisons over two structured workloads considered  above, as well as the same workloads with the order of the cell conditions permuted.

The results show that using the Fourier or the Wavelet strategy as the set of design queries \revision{doubles the error}{introduces $20\%$ more error} over all one dimensional range queries and achieves the same error on two-way marginals.  However these design queries can not maintain their performance for workloads represented under a permutation of the cell conditions: they are worse than the eigen-queries by more than \revision{an order of magnitude}{4 times} over the permuted one-dimensional range queries. 

\subsection{Experimental Conclusions} \label{sec:sub:con}

The experimental results show that, for the workloads specifically targeted by competing techniques, those techniques achieve error that is not too far from optimal (usually a factor of about \revision{1.5 to 4}{1.2 to 3.4} times the lower bound on error). But for broader classes or workloads, or ad hoc subsets of structured workloads, existing techniques are limited and the adaptivity of the Eigen-Design can improve relative or absolute error by a larger factor.  We have confirmed the versatility of our algorithm, as it improves on all competing techniques for virtually every workload considered.  \revision{}{The one exception is the highly skewed CDF workload.  The lowest error strategy we are aware of for this workload is produced by our design algorithm, but with an alternative basis.}

\vspace*{-.6ex}
\section{Related Work}\label{sec:related}
The present work uses the framework of the matrix mechanism to develop an adaptive query answering algorithm.  The original work on the matrix mechanism \cite{Li:2010Optimizing-Linear} described and analyzed in a unified framework two prior techniques specifically tailored to range queries.  The first used a wavelet transformation \cite{xiao2010differential}; the second used a hierarchical set of queries followed by inference \cite{Hay:2010Boosting-the-Accuracy}.  Originally, the matrix mechanism focused mainly on $\epsilon$-differential privacy, although $(\epsilon,\delta)$-differential privacy was also considered briefly. Prior work on the matrix mechanism never considered strategies beyond those proposed in the previous literature, or natural candidates like the identity matrix.  \revision{}{The convex optimization formalization in prior work only runs on small $n$ ($n<64$) and cannot be used in practice.}

Low order marginals are studied in \cite{barak2007privacy} using Fourier transformation.  They also consider enforcing integral consistency on the output, an objective we do not consider here.  Recently, Ding et al. proposed an adaptive algorithm to answer workloads consisting of data cube queries~\cite{Ding:2011fk}, which (described in our terms) considers strategies composed only of individual marginal queries and optimizes the workload error approximately. The algorithm adapts a known approximation algorithm for the subset-sum problem and cannot be applied to general linear queries.  Most of these techniques focus on $\epsilon$-differential privacy, however they are actually more effective under $(\epsilon,\delta)$- differential privacy, so comparisons with our algorithms are meaningful.

The error rates of the matrix mechanism are independent of the database instance.  Recently, a number of data dependent algorithms for answering linear queries under differential privacy have been proposed. Xiao et al.~\cite{xiaodifferentially} propose a method for computing a strategy matrix using KD-trees, and Cormode et al. \cite{Cormode11Differentially} propose a related method in which a differentially-private median computation is used to guide hierarchical range queries.  While promising, these approaches appear to restrict the strategy to hierarchical structures which we have shown are suboptimal for many workloads. Dynamic strategy selection can also increase computation cost.  These tradeoffs deserve further investigation.  

Focusing on relative error, Xiao et al.~\cite{Xiao11iReduct:} propose a data-dependent algorithm to minimize the relative error with an innovative resampling function.  Data-dependent interactive (as opposed to batch) mechanisms have been considered by 
Roth and Roughgarden~\cite{Roth:2010The-Median-Mechanism:}, who answer predicate queries on databases with 0-1 entries. Hardt et. al~\cite{hardt2010multiplicative} provide a linear time algorithm for the same query and database setting.  

\vspace*{-2.2ex}
\section{Conclusions and Future Work}\label{sec:conclusion}
We have described an adaptive mechanism for answering complex workloads of counting queries under differential privacy.  The mechanism can be seen to automatically select, for a given workload, a noise distribution composed of linear combinations of independent Gaussian noise.  With no reduction in privacy, the mechanism can significantly reduce error over competing techniques and is close to optimal with respect to the class of perturbation methods considered. 

In the future we hope to extend our theoretical approximation bounds to the eigen-separation and principal vector optimizations, and \revision{continue to explore the factors that make some workloads harder to answer accurately than others.}{apply our approach to non-linear queries.}

\vspace*{-2ex}
\paragraph*{Acknowledgements} Li and Miklau were supported by the NSF through grants CNS-1012748 and IIS-0964094.  We
are grateful for the comments of the anonymous reviewers. 
%
%
%

\bibliographystyle{abbrv} 
{ 
\bibliography{bib/paper} 

\begin{thebibliography}{10}

\bibitem{cvxopt}
http://abel.ee.ucla.edu/cvxopt/.

\bibitem{mmech-code}
http://dbgroup.cs.umass.edu/code/.

\bibitem{dsdp5}
http://www.mcs.anl.gov/hs/software/dsdp/.

\bibitem{barak2007privacy}
B.~Barak, K.~Chaudhuri, C.~Dwork, S.~Kale, F.~McSherry, and K.~Talwar.
\newblock Privacy, accuracy, and consistency: A holistic solution to
  contingency table release.
\newblock In {\em PODS}, pages 273--282, 2007.

\bibitem{boyd2004convex}
S.~Boyd and L.~Vandenberghe.
\newblock {\em {Convex optimization}}.
\newblock Cambridge University Press, 2004.

\bibitem{Cormode11Differentially}
G.~Cormode, M.~Procopiuc, E.~Shen, D.~Srivastava, and T.~Yu.
\newblock Differentially private spatial decompositions.
\newblock In {\em ICDE}, 2012.

\bibitem{Ding:2011fk}
B.~Ding, M.~Winslett, J.~Han, and Z.~Li.
\newblock Differentially private data cubes: optimizing noise sources and
  consistency.
\newblock In {\em SIGMOD}, pages 217--228, 2011.

\bibitem{Dwork:2011A-firm-foundation}
C.~Dwork.
\newblock A firm foundation for private data analysis.
\newblock {\em Communications of the ACM}, 54(1):86--95, 2011.

\bibitem{Dwork:2006Our-Data-Ourselves:}
C.~Dwork, K.~Kenthapadi, F.~McSherry, I.~Mironov, and M.~Naor.
\newblock Our data, ourselves: Privacy via distributed noise generation.
\newblock In {\em EUROCRYPT}, volume 4004/2006, pages 486--503, 2006.

\bibitem{Dwork:2006Calibrating-Noise}
C.~Dwork, F.~McSherry, K.~Nissim, and A.~Smith.
\newblock Calibrating noise to sensitivity in private data analysis.
\newblock In {\em TCC}, volume 3876/2006, pages 265--284, 2006.

\bibitem{ghosh2009universally}
A.~Ghosh, T.~Roughgarden, and M.~Sundararajan.
\newblock Universally utility-maximizing privacy mechanisms.
\newblock In {\em STOC}, pages 351--360, 2009.

\bibitem{hardt2010multiplicative}
M.~Hardt and G.~Rothblum.
\newblock {A multiplicative weights mechanism for privacy-preserving data
  analysis}.
\newblock In {\em FOCS}, pages 61--70, 2010.

\bibitem{Hay:2010Boosting-the-Accuracy}
M.~Hay, V.~Rastogi, G.~Miklau, and D.~Suciu.
\newblock Boosting the accuracy of differentially private histograms through
  consistency.
\newblock {\em PVLDB}, 3(1-2):1021--1032, 2010.

\bibitem{Li:2010Optimizing-Linear}
C.~Li, M.~Hay, V.~Rastogi, G.~Miklau, and A.~McGregor.
\newblock Optimizing linear counting queries under differential privacy.
\newblock In {\em PODS}, pages 123--134, 2010.

\bibitem{Li11Measuring}
C.~Li and G.~Miklau.
\newblock Measuring the achievable error of query sets under differential
  privacy.
\newblock {\em CoRR}, abs/1202.3399, 2012.

\bibitem{mcsherry2009privacy}
F.~McSherry.
\newblock Privacy integrated queries: An extensible platform for
  privacy-preserving data analysis.
\newblock In {\em SIGMOD}, pages 89--97, 2009.

\bibitem{McSherry:2009fk}
F.~McSherry and I.~Mironov.
\newblock Differentially private recommender systems: Building privacy into the
  netflix prize contenders.
\newblock In {\em SIGKDD}, pages 627--636, 2009.

\bibitem{Pukelsheim93Optimal}
F.~Pukelsheim.
\newblock {\em Optimal design of experiments}.
\newblock Wiley-Interscience, 1993.

\bibitem{Roth:2010The-Median-Mechanism:}
A.~Roth and T.~Roughgarden.
\newblock Interactive privacy via the median mechanism.
\newblock In {\em STOC}, pages 765--774, 2010.

\bibitem{Xiao11iReduct:}
X.~Xiao, G.~Bender, M.~Hay, and J.~Gehrke.
\newblock {iReduct}: Differential privacy with reduced relative errors.
\newblock In {\em SIGMOD}, pages 229--240, 2011.

\bibitem{xiao2010differential}
X.~Xiao, G.~Wang, and J.~Gehrke.
\newblock Differential privacy via wavelet transforms.
\newblock In {\em ICDE}, pages 1200--1214, 2010.

\bibitem{xiaodifferentially}
Y.~Xiao, L.~Xiong, and C.~Yuan.
\newblock Differentially private data release through multidimensional
  partitioning.
\newblock In {\em SDM}, pages 150--168, 2010.

\end{thebibliography}
\normalsize
}

\end{document}